\documentclass[%
 reprint,
 amsmath,amssymb,
 aps,
 pra,
]{revtex4-1}

\usepackage{graphicx}
\usepackage{dcolumn}
\usepackage{bm}

\usepackage{braket} 

\begin{document}

\preprint{APS/123-QED}

\title{Semiclassical Analysis of the Wigner $9J$-Symbol \\ with Small and Large Angular Momenta}

\author{Liang Yu}
\email{liangyu@wigner.berkeley.edu}

\author{Robert G. Littlejohn}

\affiliation{
 Department of Physics, University of California, Berkeley, California 94720 USA \\
}

\date{\today}

\begin{abstract}

We derive a new asymptotic formula for the Wigner $9j$-symbol, in the limit of one small and eight large angular momenta, using a novel gauge-invariant factorization for the asymptotic solution of a set of coupled wave equations. Our factorization eliminates the geometric phases completely, using gauge-invariant non-canonical coordinates, parallel transports of spinors, and quantum rotation matrices. Our derivation generalizes to higher $3nj$-symbols. We display without proof some new asymptotic formulas for the $12j$-symbol and the $15j$-symbol in the appendices. This work contributes a new asymptotic formula of the Wigner $9j$-symbol to the quantum theory of angular momentum, and serves as an example of a new general method for deriving asymptotic formulas for $3nj$-symbols.

\begin{description}
\item[PACS numbers] 03.65.Sq, 02.30.Ik, 03.65.Vf
\end{description}
\end{abstract}

                             
\maketitle

\section{INTRODUCTION}
 
This paper contains new asymptotic formulas for the $9j$-, $12j$-, and $15j$-symbols when some quantum numbers are large and others small.  In particular, we derive in full detail an asymptotic formula for the $9j$-symbol in the limit of one small and eight large angular momenta, as well as several other, similar results for the $9j$- $12j$- and $15j$-symbols.  The main theoretical tool we use is a generalization of the Born-Oppenheimer approximation, in which the small angular momenta are the fast degrees of freedom and the large angular momenta are the slow degrees of freedom. 

In standard applications in molecular physics, the Born-Oppenheimer
approximation couples a set of electronic modes by means of a
potential energy matrix, in which the matrix elements are all
functions of position and so commute with one another.  In our
applications the modes of the fast degrees of freedom are coupled by a
matrix of noncommuting operators.  The required generalization of the
Born-Oppenheimer approximation in this case was presented by
Littlejohn and Flynn \cite{littlejohn1991} and applied by those authors
to a semiclassical treatment of spin-orbit
coupling \cite{littlejohn1992}.  The method involves diagonalizing the
matrix of operators by means of the Moyal star product, in a
pertubation expansion in powers of $\hbar$.  The results are
interesting geometrically, in that there appears a bundle whose base
space is the classical phase space of the slow degrees of freedom,
with a fiber that is the Hilbert space for the fast degrees of
freedom.  The fiber bundle carries a Berry's connection, whose
curvature contributes to the symplectic form on the base space.  This
paper makes extensive use of the techniques developed in
Refs.~\cite{littlejohn1991,littlejohn1992}, and assumes a familiarity with
them.  In addition we note the work of Emmrich and
Weinstein \cite{weinstein1996}, which reinterpreted the transformations
of Littlejohn and Flynn as a deformation of the Weyl symbol
correspondence and placed the whole procedure in a more mathematical
setting.

In this paper, as in the spin-orbit problem \cite{littlejohn1992}, the
small angular momentum is represented by exact linear algebra, and the
states are represented by multicomponent wave-functions.  We use
Schwinger's model \cite{schwinger1952} to represent the large angular
momenta.  Some of the issues we face in this paper are non-Abelian
symmetry groups in the presence of the Born-Oppenheimer approximation
and the problem of constructing gauge-invariant expressions for wave
functions and matrix elements.  

Part of the significance of this work is a new asymptotic formula for
the Wigner $9j$-symbol in terms of a modified form of the well-known
Ponzano-Regge formula \cite{ponzano-regge-1968} for the $6j$-symbol.  The
other part is the demonstration of a new technique that is generally
applicable to the studies of higher $3nj$-symbols.

We will briefly review some of the previous works on the asymptotics of the $3nj$-symbol, when some of the angular momenta are small and others are large. Basic references for the definitions and properties of the $3nj$-symbols include Edmonds \cite{edmonds1960}, Biedenharn and Louck \cite{biedenharn1981}, Brink and Satchler \cite{brink1968}, and Varshalovich \cite{varshalovich1981}. Let us use the notation $(s, l)$ to classify the various asymptotic limits of the $3nj$-symbols, where $s$ and $l$ denote the number of small and large angular momenta, respectively. For a given $3nj$-symbol, not all possibilities $l + s = 3nj$ are allowed, since the sum of two small angular momenta must be small. For the $3j$-symbol, in 1957, Brussaard and Tolhoek \cite{brussaard1957} used the Stirling's approximation in a sum formula for the Clebsch-Gordan coefficient to derive the $(1,2)$ case. In 1968, Ponzano and Regge \cite{ponzano-regge-1968} used intuitive methods to guess the $(0,3)$ case for the $3j$-symbol, as well as the $(0, 6)$ case for the $6j$-symbol. in 1960, Edmonds \cite{edmonds1960} found the $(1,5)$ case, which trivially leads to the $(2, 4)$ case \cite{Nikiforov1991}.  In 1999, Watson \cite{watson1999b, watson1999a} derived the $(3,3)$ case for the $6j$-symbol, as well as the $(4, 5)$ case for the $9j$-symbol. More recently, Anderson et. al. \cite{anderson2008, anderson2009} extended Watson's result of the $9j$-symbol to the $(6,3)$ case where all the small angular momenta are in different rows. The other $(6,3)$ case where all three small angular momenta are in the same row is still an open problem. Later Haggard and Littlejohn \cite{littlejohn2010b} found the formula for $(0, 9)$ case. Our results in this paper will fill in most of the gap for the $9j$-symbol by deriving asymptotic formulas of the $(1,8)$ case and the $(2,7)$ case.

Now we give a brief overview of our derivation. After we apply the techniques from \cite{littlejohn1992} to the $9j$-symbol, each multicomponent wave-function consists of a spinor factor and a factor in the form of a scalar WKB solution. The action in the scalar WKB solution is the integral of $p \, dx$ on a Lagrangian manifold. We introduce a nearby Lagrangian manifold, the ``$6j$ manifold,'' which is just the $A$- or $B$-manifold from \cite{littlejohn2010}. This Lagrangian manifold appears in the semiclassical analysis of the $6j$-symbol, and its action is analyzed in \cite{littlejohn2010}. We perturb the unknown action around the action on the $6j$ manifold, through a two steps process. First we introduce an intermediate manifold using the coordinate transformation that gives rise to the gauge invariant coordinates from \cite{littlejohn1992}. By writing the initial gauge-dependent action as a line integral, and perturbing the path for the line integral from the initial Lagrangian manifold to a path on the intermediate manifold, we can write the original line integral as a sum of a gauge-invariant line integral and a geometric phase. Then we perturb the gauge-invariant line integral again by following nearby Hamiltonian flows on the intermediate manifold and on the $6j$ manifold, respectively. This second perturbation further splits the gauge-invariant line integral into a known action on the $6j$ manifold, and a perturbation. When we combine the geometric phase and the perturbation with the spinor field on the Lagrangian manifold, we obtain a field of quantum rotations acting on a fixed spinor. As a result, we obtain a factorization of the multicomponent wave function into a known scalar WKB factor, namely, those for the $6j$-symbol from \cite{littlejohn2010}, and a field of quantum rotations of a reference spinor. Although the quantum rotations are difficult to handle by themselves, they partially cancel out when we take an inner product between two multicomponent wave functions. In the case of the two $6j$ manifolds, there are two components in the intersection set, and only one rotation is required for each Lagrangian manifold. It turns out that the spinor inner product generates a factor of a Wigner $d$-matrix, and an additional relative phase between the two terms of the WKB approximation. In the end, we find an asymptotic formula for the $9j$-symbol in terms of a Wigner $d$-matrix and a modified Ponzano-Regge formula for a $6j$-symbol.

Now we give an outline of the paper. In Sec. \ref{SLQN9j: sec_SLQN9j: eq_main_formula}, we display the main result of this paper, namely, a new asymptotic formula for the $9j$-symbol where one of the angular momenta is small. In Sec. \ref{SLQN9j: sec_9j_model}, we describe the representation we use for the $9j$-symbol, and express it as an inner product between two multicomponent wave-functions. In Sec. \ref{SLQN9j: sec_factorization}, we derive a gauge invariant factorization of the wave-functions. In Sec. \ref{SLQN9j: sec_9j_derivation}, we use this factorization to derive the asymptotic formula for the $9j$-symbol. Some plots of our formula are presented in Sec. \ref{SLQN9j: sec_numeric}. The last section contains comments and conclusions. The appendices contain some generalized results for the higher $3nj$-symbols.

\section{\label{SLQN9j: sec_SLQN9j: eq_main_formula}  AN ASYMPTOTIC FORMULA FOR THE $9j$-SYMBOL}

We quote the main result of this paper, namely, a new asymptotic formula for the $9j$-symbol, where one angular momentum, $j_3 = s $, is small compared to the others. We have chosen $j_3$ to be the small quantum number in the main formula below, but any other choice can be reduced to this case by the symmetries of the $9j$-symbol. The formula is

\begin{figure}[tbhp]
\begin{center}
\includegraphics[width=0.30\textwidth]{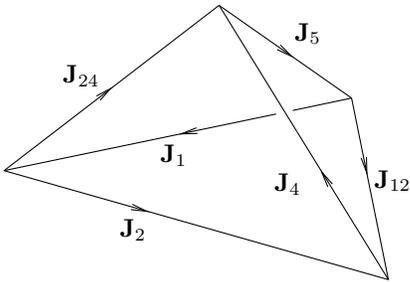}
\caption{The tetrahedron constructed from the six edge lengths $J_r$, $r = 1,2,4,5,12,24$. }
\label{SLQN9j: fig_tetrahedra}
\end{center}
\end{figure}

\begin{figure}[tbhp]
\begin{center}
\includegraphics[width=0.36\textwidth]{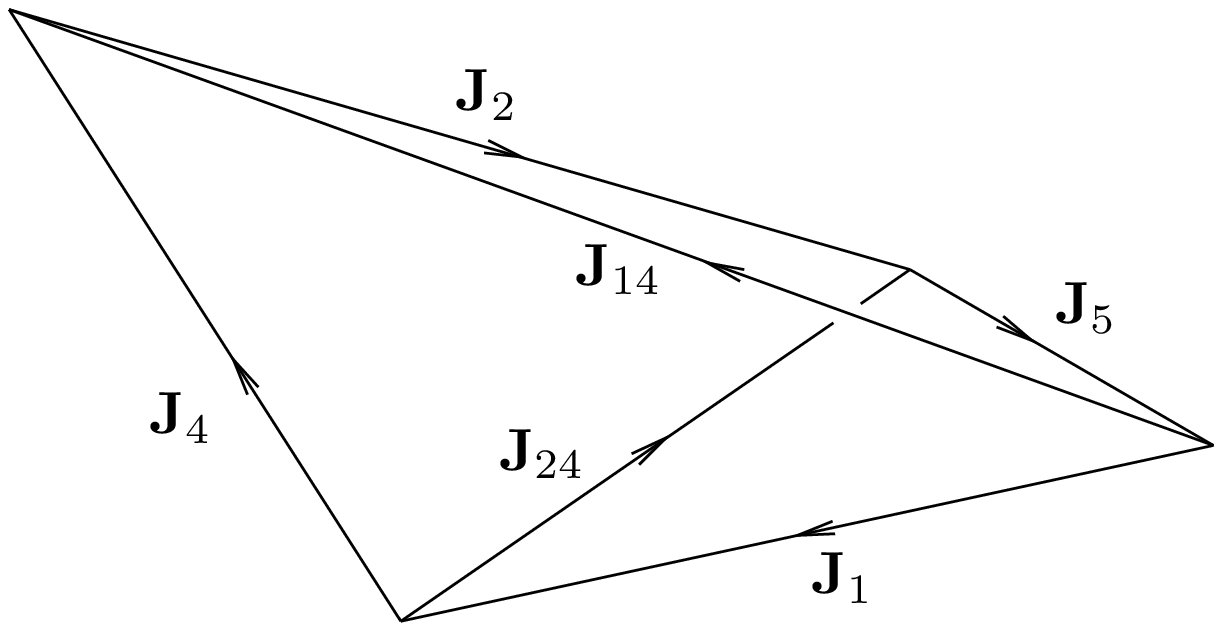}
\caption{The vectors configuration from Fig.\ \ref{SLQN9j: fig_tetrahedra} are rearranged from ${\mathbf J}_1+{\mathbf J}_2+{\mathbf J}_4+{\mathbf J}_5={\mathbf 0}$ to ${\mathbf J}_1+{\mathbf J}_4+{\mathbf J}_2+{\mathbf J}_5={\mathbf 0}$. The intermediate vector ${\mathbf J}_{12}$ no longer appears. Instead, we have ${\mathbf J}_{14} = {\mathbf J}_1 + {\mathbf J}_4$.}
\label{SLQN9j: fig_tetrahedra_1p}
\end{center}
\end{figure}

\begin{figure}[tbhp]
\begin{center}
\includegraphics[width=0.35\textwidth]{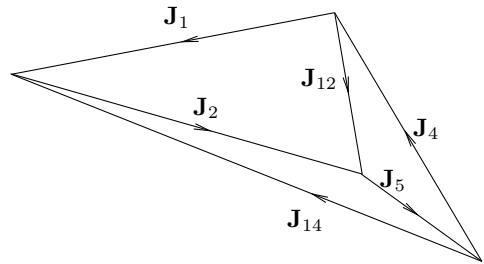}
\caption{The vectors configuration from Fig.\ \ref{SLQN9j: fig_tetrahedra} are rearranged from ${\mathbf J}_1+{\mathbf J}_2+{\mathbf J}_4+{\mathbf J}_5={\mathbf 0}$ to ${\mathbf J}_4+{\mathbf J}_1+{\mathbf J}_2+{\mathbf J}_5={\mathbf 0}$. The intermediate vector ${\mathbf J}_{24}$ no longer appears. Instead, we have ${\mathbf J}_{14} = {\mathbf J}_1 + {\mathbf J}_4$. }
\label{SLQN9j: fig_tetrahedra_4p}
\end{center}
\end{figure}

\begin{widetext}
\begin{equation}
\left\{
  \begin{array}{ccc}
    j_1 & j_2 & j_{12} \\ 
    s & j_4 & j_{34} \\ 
    j_{13} & j_{24} & j_5 \\
  \end{array} 
  \right\}
	= \frac{(-1)^{j_1+j_2+j_4+j_5+2 s +\nu}}{\sqrt{(2j_{13}+1) (2j_{34}+1)  \, (12 \pi V)}} \, 
	\cos \left(  \sum_i  \, (j_i+\frac{1}{2}) \, \psi_i  + \frac{\pi}{4} - s \pi + \mu \phi_1 +\nu \phi_4  \right) 
	\, d^{s}_{\nu \, \mu} (\theta) \, . 
\label{SLQN9j: eq_main_formula}
\end{equation}
\end{widetext}
Here the indices on the $d$-matrix are given by $\mu = j_{13}-j_1$ and $\nu = j_{34}-j_4$. They are of the same order of magnitude as the small parameter $s$. The sum in the argument of the cosine runs over the six large angular momenta $i = 1, 2, 4, 5, 12, 24$. Out of the eight large angular momenta, these are the ones that do not involve the index $i=3$. The geometric quantities $V$, $\psi_i$, $\phi_1$, $\phi_4$, and $\theta$ are functions of the vector configuration of the tetrahedron in Fig.\ \ref{SLQN9j: fig_tetrahedra}, the construction of which we will describe shortly. In exactly the same way as they appear in the Ponzano Regge formula \cite{ponzano-regge-1968}, $V$ is the volume of the tetrahedron, and each $\psi_i$ is the external dihedral angle at the edge $J_i$, where $\psi \in [0, \pi]$. The angle $\phi_1 \in [0, \pi]$ is an interior dihedral angle of a tetrahedron containing the edge $J_1$, but the tetrahedron in question is not the one in Fig.\ \ref{SLQN9j: fig_tetrahedra}, but rather a related, second tetrahedron in Fig.\ \ref{SLQN9j: fig_tetrahedra_1p}. Similarly, the angle $\phi_4 \in [0, \pi]$ is the internal dihedral angle at $J_4$ in a third tetrahedron in Fig.\ \ref{SLQN9j: fig_tetrahedra_4p}. The angle $\theta \in [0, \pi]$ is not a dihedral angle, but it is the angle between ${\mathbf J}_1$ and ${\mathbf J}_4$, which is adjacent in Fig.\ \ref{SLQN9j: fig_tetrahedra_1p} and \ref{SLQN9j: fig_tetrahedra_4p}. Explicitly, the angles $\phi_1$, $\phi_4$, and $\theta$ are defined in terms of the vectors as follows:

\begin{equation}
\label{SLQN9j: eq_phi1_def}
\phi_1 = \pi -  \cos^{-1} \left( \frac{ ( {\mathbf J}_1 \times {\mathbf J}_4 ) \cdot  ( {\mathbf J}_1 \times {\mathbf J}_5 ) }{ | {\mathbf J}_1 \times {\mathbf J}_4 | \,  | {\mathbf J}_1 \times {\mathbf J}_5 | }  \right) \, , 
\end{equation}

\begin{equation}
\phi_4 = \pi -  \cos^{-1} \left( \frac{ ( {\mathbf J}_4 \times {\mathbf J}_1 ) \cdot  ( {\mathbf J}_4 \times {\mathbf J}_5 ) }{ | {\mathbf J}_4 \times {\mathbf J}_1 | \,  | {\mathbf J}_4 \times {\mathbf J}_5 | }  \right) \, , 
\end{equation}

\begin{equation}
\label{SLQN9j: eq_theta_def}
\theta =  \cos^{-1} \left( \frac{{\mathbf J}_1 \cdot {\mathbf J}_4}{J_1 \, J_4} \right) =  \cos^{-1} \left( \frac{{\mathbf J}_{12} \cdot {\mathbf J}_4 - {\mathbf J}_{2} \cdot {\mathbf J}_4 }{J_1 \, J_4}  \right) \, .
\end{equation}

To use Eq.\ (\ref{SLQN9j: eq_main_formula}), we construct the vector configurations of the tetrahedron in Fig.\ \ref{SLQN9j: fig_tetrahedra}, given the six edge lengths $J_r$, $r = 1,2,4,5,12,24$, according to the procedure in the Appendix of \cite{littlejohn2009}. There, a Gram matrix $G$, which consists of dot products between three vectors at a vertex of the tetrahedron in Fig.\ \ref{SLQN9j: fig_tetrahedra}, say ${\mathbf J}_1$, ${\mathbf J}_{12}$, $-{\mathbf J}_5$, is expressed in terms of the six edge lengths $J_r = j_r + 1/2$, $r = 1,2,4,5,12,24$. Then a singular decomposition of $G$ gives a vector configuration of the three vectors, and the remaining three vectors are constructed from the different pairs of the first three vectors. For instance, the resulting vector configuration for the values $(J_1, J_2, J_4, J_5, J_{12}, J_{24}) = (101, 123, 88, 64.5, 68.5, 92.5)$ is illustrated in Fig.\ \ref{SLQN9j: fig_tetrahedra}. Notice the tetrahedra from Fig.\  \ref{SLQN9j: fig_tetrahedra_1p} and Fig.\ \ref{SLQN9j: fig_tetrahedra_4p} contain the edge length $J_{14}$, which is not related to any of the $9$ quantum numbers of the $9j$-symbol in Eq.\  (\ref{SLQN9j: eq_main_formula}), so we cannot use the same procedure above directly to construct their vector configurations. However, the vectors ${\mathbf J}_r$, $r = 1,2,4,5$ in the tetrahedra in Fig.\ \ref{SLQN9j: fig_tetrahedra_1p} and Fig.\ \ref{SLQN9j: fig_tetrahedra_4p} are related to those in Fig.\ \ref{SLQN9j: fig_tetrahedra} by parallel translations, and ${\mathbf J}_{14} = {\mathbf J}_1 + {\mathbf J}_4$, so we can find the vector configurations in Fig.\ \ref{SLQN9j: fig_tetrahedra_1p} and Fig.\ \ref{SLQN9j: fig_tetrahedra_4p} once we have those in Fig.\ \ref{SLQN9j: fig_tetrahedra}. Thus, the quantities $V, \psi_i, \phi_1, \phi_4, \theta$ are all functions of the six edge lengths $J_r$, $r =1,2,4,5,12,24$.

\section{\label{SLQN9j: sec_9j_model}  A REPRESENTATION FOR THE $9j$-SYMBOL}

\subsection{The $9j$-symbol as a scalar product}

We express the $9j$-symbol as an inner product of two multicomponent wave-functions, using a variation of Edmond's definition, (6.4.2) in \cite{edmonds1960}, as follows, 

\begin{eqnarray}
 && \left\{
   \begin{array}{ccc}
    j_1 & j_2 & j_{12} \\ 
    s & j_4 & j_{34} \\ 
    j_{13} & j_{24} & j_5 \\
  \end{array} 
  \right \}    \\  \nonumber
&=& \frac{\braket{ b  |  a } }{[(2j_{12}+1)(2j_{34}+1)(2j_{13}+1)(2j_{24}+1)]^{\frac{1}{2}}} \, . 
\end{eqnarray}
Here $\ket{a}$ and $\ket{b}$ are normalized simultaneous eigenstates of lists of operators with certain eigenvalues. We will ignore the phase conventions of $\ket{a}$ and $\ket{b}$ for now, since we did not use them to derive our formula. In our notation, the two states are 
	
\begin{equation}
\label{SLQN9j: eq_a_state}
\ket{a} =  \left| 
\begin{array} { @{\;}c@{\,}c@{\;}c@{\;}c@{\,}c@{\;}c@{\;}c@{\;}c@{}}
	\hat{I}_1 & \hat{I}_2 & {\mathbf S}^2 & \hat{I}_4 & \hat{I}_5 & \hat{\mathbf J}_{13}^2 & \hat{\mathbf J}_{24}^2 & \hat{\mathbf J}_{\text{tot}}  \\
	j_1 & j_2 & s & j_4 & j_5 & j_{13} & j_{24} & {\mathbf 0} 
\end{array}  \right>  \,  , 
\end{equation}

\begin{equation}
\label{SLQN9j: eq_b_state}
\ket{b} =  \left| 
\begin{array} { @{\;}c@{\,}c@{\;}c@{\;}c@{\,}c@{\;}c@{\;}c@{\;}c@{}}
	\hat{I}_1 & \hat{I}_2 & {\mathbf S}^2 & \hat{I}_4 & \hat{I}_5 & \hat{\mathbf J}_{12}^2 & \hat{\mathbf J}_{34}^2 & \hat{\mathbf J}_{\text{tot}}  \\
	j_1 & j_2 & s & j_4 & j_5 & j_{12} & j_{34} & {\mathbf 0} 
\end{array}  \right>  \, . 
\end{equation}
In the above notation, the large ket lists the operators on the top row, where a hat is used to distinguish an operator from its classical symbol function, and the corresponding quantum numbers are listed on the bottom row.  
 
Now we specify the Hilbert space where the two states $\ket{a}$ and $\ket{b}$ live. The states belong to a total Hilbert space of five angular momenta ${\cal H}_1 \otimes {\cal H}_2 \otimes {\cal H}_4 \otimes {\cal H}_5  \otimes {\cal H}_s $. Each large angular momentum $\hat{\mathbf J}_r$, $r = 1,2,4,5$, acts on its own copy of the Schwinger space of two harmonic oscillators, namely, ${\cal H}_r = L^2({\mathbb R}^2)$ \cite{schwinger1952}. The small angular momentum ${\mathbf S}$ is represented by the usual $2s+1$ dimensional representation of $SU(2)$, that is to say, it acts on ${\cal H}_s = {\mathbb C}^{2s + 1}$. 

We now define the lists of operators in Eq.\  (\ref{SLQN9j: eq_a_state}) and Eq.\  (\ref{SLQN9j: eq_b_state}). First we look at the operators $\hat{I}_r$, $r = 1,2,4,5$, $\hat{\mathbf J}_{12}^2$, and $\hat{\mathbf J}_{24}^2$, which act only on the large angular momentum spaces, ${\cal H}_r = L^2({\mathbb R}^2) $, $r = 1,2,4,5$, each of which can be viewed as a space of wave-functions $\psi(x_{r1}, \, x_{r2})$ for two harmonic oscillators of unit frequency and mass. Let $\hat{a}_{r\mu} = (\hat{x}_{r\mu} + i \hat{p}_{r\mu})/\sqrt{2}$, and $\hat{a}_{r\mu}^\dagger = (\hat{x}_{r\mu} - i \hat{p}_{r\mu})/\sqrt{2}$, $\mu = 1,2$, be the usual annihilation and creation operators. The operators $\hat{I}_r$ and $\hat{J}_{ri}$ are constructed from  $\hat{a}$ and $\hat{a}^\dagger$ as follows, 

\begin{equation}
\hat{I}_r = \frac{1}{2} \, \sum_\mu \hat{a}_{r\mu}^\dagger \hat{a}_{r\mu} \, ,  \quad \quad \hat{J}_{ri} = \frac{1}{2} \, \sum_{\mu\nu} \hat{a}^\dagger_{r \mu} (\sigma_i)_{\mu\nu} \hat{a}_{r\nu}  \, ,
\end{equation}
where $i = 1,2,3$, and $\sigma_i$ are the Pauli matrices. The quantum numbers $j_r$, $r = 1,2,4,5$, specify the eigenvalues of both $\hat{I}_r$ and $\hat{\mathbf J}_r^2$ to be $j_r$ and $j_r(j_r + 1)$, respectively.

The operators $\hat{\mathbf J}_{12}^2$ and $\hat{\mathbf J}_{24}^2$ that define the intermediate couplings of the large angular momenta are defined as partial sums of the $\hat{\mathbf J}_r$, 

\begin{equation}
\label{SLQN9j: eq_j12_j24_vectors}
	\hat{\mathbf J}_{12} = \hat{\mathbf J}_1 + \hat{\mathbf J}_2 \, ,   \quad \quad  \hat{\mathbf J}_{24} = \hat{\mathbf J}_2 + \hat{\mathbf J}_4  \, .
\end{equation}
The quantum numbers $j_{12}$ and $j_{24}$ specify the eigenvalues of the operators $\hat{\mathbf J}_{12}^2$ and $\hat{\mathbf J}_{24}^2$, that is, $j_{12} (j_{12}+1)$ and $j_{24}(j_{24}+1)$, respectively. See  \cite{littlejohn2007} for more detail on the Schwinger model.

Now we turn our attention to the operator $S^2$ that acts only on the small angular momentum space ${\mathbb C}^{2s+1}$. Let ${\mathbf S}$ be the vector of dimensionless spin operators represented by $(2s+1) \times (2s+1)$ matrices that satisfy the $SU(2)$ commutation relations

\begin{equation}
[S_i, S_j] = i \, \sum_k \, \epsilon_{ijk} \, S_k \, .
\end{equation}
The Casimir operator, ${\mathbf S}^2 = s (s+1)$, is proportional to the identity operator, so its eigenvalue equation is trivially satisfied.

The remaining operators $\hat{\mathbf J}_{13}^2$, $\hat{\mathbf J}_{34}^2$, and $\hat{\mathbf J}_{\text{tot}}$ act on both the ${\mathcal H}_r$ and ${\mathcal H}_s$. As in the spin-orbit problem \cite{littlejohn1992}, these are represented by nondiagonal matrices of differential operators. Even for the operators $\hat{I}_r$, $\hat{J}_{12}^2$, $\hat{J}_{24}$, above, they should be viewed as differential operators times the identity matrix. Similarly, the operators ${\mathbf S}$, $S^2$ should be viewed as matrices times the identity operator on all ${\mathcal H}_r$. The following equations define $\hat{\mathbf J}_{13}^2$, $\hat{\mathbf J}_{34}^2$, and $\hat{\mathbf J}_{\text{tot}}$. 

\begin{eqnarray}
\label{SLQN9j: eq_j13_square}
({\hat{J}}_{13}^2)_{\alpha \beta} &=& [  \hat{I}_1 (\hat{I}_1+1) + \hbar^2 s(s+1) ] \delta_{\alpha \beta} + 2  \hbar \, \hat{\mathbf J}_1 \cdot {\mathbf S}_{\alpha \beta} , \quad \quad  \\
\label{J34_sq}
({\hat{J}}_{34}^2)_{\alpha \beta}  &=& [  \hat{I}_4 (\hat{I}_4+1) + \hbar^2 s(s+1) ] \delta_{\alpha \beta} + 2 \hbar \, \hat{\mathbf J}_4 \cdot {\mathbf S}_{\alpha \beta} ,   \\
\label{SLQN9j: eq_total_J_vector}
(\hat{\mathbf J}_{\text{tot}})_{\alpha \beta}  &=& ( \hat{\mathbf J}_1 + \hat{\mathbf J}_2 + \hat{\mathbf J}_4 + \hat{\mathbf J}_5 ) \delta_{\alpha \beta} + \hbar \, {\mathbf S}_{\alpha \beta} .
\end{eqnarray}

\subsection{The multicomponent wave-functions}

We follow the approach used in \cite{littlejohn1991, littlejohn1992} to find the WKB form of the multicomponent wave-functions $\psi^a_\alpha(x) = \braket{x, \alpha | a}$ and $\psi^b_\alpha(x) = \braket{x, \alpha | b}$. Let us focus on $\psi^a_\alpha(x)$, since the treatment of $\psi^b$ is analogous. Temporarily, we will drop the $a$ superscript.

Let $\hat{D}_i$, $i = 1, \dots, 10$ denote the the operators listed in the definition of $\ket{a}$ in Eq.\ (\ref{SLQN9j: eq_a_state}). We seek a unitary operator $\hat{U}$, such that $\hat{D}_i$ for all $i=1, \dots, 10$ are diagonalized when conjugated by $\hat{U}$, that is,   

\begin{equation}
 \sum_{\alpha \beta} \, \hat{U}^\dagger_{\alpha \, \mu} (\hat{D}_i)_{\alpha \, \beta} \, \hat{U}_{\beta \, \nu} = (\hat{\Lambda}_i)_{\mu \, \nu} \, , 
\end{equation}
where $\hat{\Lambda}_i$, $i =1, \dots, 10$ is a list of diagonal matrix operators, and where we have employed the index summation notation. Let $\phi^{(\mu)}$ be the simultaneous eigenfunction for the $\mu^{\text{th}}$ diagonal entries $\hat{\lambda}_i$ of the operators $\hat{\Lambda}_i$, then we obtain a simultaneous eigenfunction $\psi_\alpha^{(\mu)}$ of the original list of operators $\hat{D}_i$ from

\begin{equation}
\psi_\alpha^{(\mu)} = \hat{U}_{\alpha \, \mu} \, \phi^{(\mu)} \, . 
\end{equation}
Here the index $\mu$ is called the polarization index, in the same sense used in \cite{littlejohn1992}. There is no summation over the polarization index. Since we are interested in $\psi_\alpha$ only to first order in $\hbar$, all we need are the zeroth order term $U$ of the Weyl symbol matrix of $\hat{U}$ and the first order symbol matrix $\Lambda_i$ of $\hat{\Lambda}_i$. The resulting asymptotic form of the wavefunction $\psi(x)$, ( (4.6) in \cite{littlejohn1991}), is a product of a scalar WKB part $B e^{iS}$ and a spinor part $\tau$, that is,

\begin{equation}
\label{SLQN9j: eq_general_wavefunction}
\psi_\alpha^{(\mu)} (x) =  B(x) \, e^{i \, S(x) / \hbar}  \, \tau_{\alpha}^{(\mu)} (x, p) \, . 
\end{equation}
Here the action $S(x)$ and the amplitude $B(x)$ are simultaneous solutions to the Hamilton-Jacobi and the transport equations, respectively, that are associated with the Hamiltonians $\lambda^{(\mu)}_i$. The spinor $\tau^{\mu}$ is the $\mu^{\text{th}}$ column of the matrix $U$, 

\begin{equation}
\label{SLQN9j: eq_U_and_tau}
\tau_{\alpha}^{(\mu)} (x, p) = U_{\alpha \mu} (x, p) \, , 
\end{equation}
and where $p = \partial S(x) / \partial x$. 

Now let us apply the above strategy to our problem. The Weyl symbols of the operators  $\hat{I}_r$ and $\hat{J}_{ri}$ are $I_r - 1/2$ and $J_{ri}$, respectively, where 

\begin{equation}
\label{SLQN9j: eq_I_Ji_symbols}
I_r = \frac{1}{2} \,  \sum_\mu \overline{z}_{r\mu} z_{r\mu} \, ,   \quad \quad J_{ri} = \frac{1}{2} \, \sum_{\mu\nu} \overline{z}_{r\mu} (\sigma_i)_{\mu\nu} z_{r\nu} \, ,
\end{equation}
and where  $z_{r\mu} = x_{r\mu} + ip_{r\mu}$ and $\overline{z}_{r\mu} = x_{r\mu} - ip_{r\mu}$ are the symbols of $\hat{a}$ and $\hat{a}^\dagger$, respectively. The symbols of the remaining operators have the same expressions as Eq.\  (\ref{SLQN9j: eq_j12_j24_vectors}), (\ref{SLQN9j: eq_j13_square})-(\ref{SLQN9j: eq_total_J_vector}), but without the hats.

Among the operators $\hat{D}_i$, $\hat{J}_{13}^2$ and the vector of the three operators $\hat{\mathbf J}_{\rm tot}$ are non-diagonal. First we look at the expression for $\hat{J}_{13}^2$,  Eq.\  (\ref{SLQN9j: eq_j13_square}), and see that the zeroth order term of the symbol matrix $J_{13}^2$ is already proportional to the identity matrix, so the spinor  $\tau$ must be an eigenvector for the first order term ${\mathbf J}_1 \cdot {\mathbf S}$. Let $\tau^{(\mu)}({\mathbf J}_1)$ be the eigenvector of the matrix ${\mathbf J}_1 \cdot {\mathbf S}$ with eigenvalue $\mu I_1$, that is, it satisfies

\begin{equation}
\label{SLQN9j: eq_JS_eigenvector}
({\mathbf J}_1 \cdot {\mathbf S})_{\alpha \beta} \, \tau^{(\mu)}_\beta = \mu I_1 \, \tau^{(\mu)}_\beta \, ,
\end{equation}
where $\mu = -s,\, \dots\, , \,+s$. In order to preserve the diagonal symbol matrices $I_r$ through the unitary transformation, we must choose the spinor $\tau^{(\mu)}$ to depend only on the direction of ${\mathbf J}_1$. One possible choice of $\tau^{(\mu)}$ is the ``north standard gauge'', (see Appendix A of \cite{littlejohn1992}), in which the spinor $\delta_{\alpha\, \mu}$ is rotated along a great circle from the $z$-axis to the direction of ${\mathbf J}_1$. Explicitly, 

\begin{equation}
\label{SLQN9j: eq_tau_NSG}
\tau^{(\mu)}_\alpha ({\mathbf J}_1) = e^{i (\mu - \alpha) \phi_1} \, d^{(s)}_{\alpha \, \mu} (\theta_1) \, , 
\end{equation}
where $(\theta_1, \phi_1)$ are the spherical coordinates that specify the direction of ${\mathbf J}_1$.  Note that this is not the only choice, since  Eq.\  (\ref{SLQN9j: eq_JS_eigenvector}) is invariant under a local $U(1)$ gauge transformations.  In other words, any other spinor $\tau' = e^{i g({\mathbf J}_1)} \, \tau$ that is related to $\tau$ by a $U(1)$ gauge transformation satisfies  Eq.\  (\ref{SLQN9j: eq_JS_eigenvector}).  This local gauge freedom is parametrized by the vector potential,

\begin{equation}
\label{SLQN9j: eq_gauge_pot_A}
{\mathbf A}^{(\mu)}_1 = i (\tau^{(\mu)} )^\dagger \, \frac{\partial \tau^{(\mu)}}{\partial {\mathbf J}_1} \, , 
\end{equation}
which transforms as ${\mathbf A}^{(\mu)'} = {\mathbf A}^{(\mu)} - \nabla_{{\mathbf J}_1} (g)$ under a local gauge transformation. Moreover, the gradient of the spinor can be expressed in terms of the vector potential, (see (A.22) in \cite{littlejohn1992}), as follows, 

\begin{equation}
\label{SLQN9j: eq_ tau_derivative}
\frac{\partial \tau^{(\mu)}}{\partial {\mathbf J}_1} = i \left( - {\mathbf A}_1^{(\mu)} + \frac{{\mathbf J}_1 \times {\mathbf S}}{I_1^2} \right) \, \tau^{(\mu)} \, .
\end{equation}
Once we obtain the complete set of spinors $\tau^{(\mu)}$, $\mu = -s, \dots, s$, we can construct the zeroth order symbol matrix $U$ of the unitary transformation from  Eq.\  (\ref{SLQN9j: eq_U_and_tau}). 

Now let us show that all the transformed symbol matrices of the operators in  Eq.\  (\ref{SLQN9j: eq_a_state}),  namely, the $\Lambda_i$, are diagonal to first order. Let us write $\hat{\Lambda} [ \hat{D} ]$ to denote the operator $\hat{U}^\dagger \hat{D} \hat{U}$, and write $\Lambda [ \hat{D} ]$ for its Weyl symbol. First, consider the operators $\hat{I}_r$, $r = 1,\dots, 4$, which are proportional to the identity matrix. Using the operator identity

\begin{equation}
[\hat{\Lambda}(\hat{I}_r)]_{\mu\nu} = \hat{U}^\dagger_{\alpha \mu} ( \hat{I}_r  \delta_{\alpha \beta} ) \hat{U}_{\beta \nu} = \hat{I}_r \delta_{\mu \nu} - \hat{U}^\dagger_{\alpha \mu} [ \hat{U}_{\alpha \nu} , \, \hat{I}_r] \, ,
\end{equation}
we find

\begin{equation}
\label{SLQN9j: eq_symbol_trick}
[\Lambda(\hat{I}_r)]_{\mu\nu}  = (I_r - 1/2) \delta_{\mu \nu} - i \hbar  U_{0\alpha \mu}^* \, \{ U_{0 \alpha \nu} , \, I_r \} \, , 
\end{equation}
where we have used the fact that the symbol of a commutator is a Poisson bracket. Since $U_{\alpha \mu} = \tau^{(\mu)}_\alpha$ is a function only of ${\mathbf J}_1$, and since the Poisson brackets $\{ {\mathbf J}_1, I_r \} = 0$ vanish for all $r = 1,2,4,5$, the second term in  Eq.\  (\ref{SLQN9j: eq_symbol_trick}) vanishes. We have

\begin{equation}
[\Lambda(\hat{I}_r)]_{\mu\nu} = (I_r - 1/2) \, \delta_{\mu\nu} \, .
\end{equation}
Similarly, we find

\begin{equation}
[\Lambda(\hat{J}_{24}^2)]_{\mu\nu} = J_{24}^2 \, \delta_{\mu\nu} \, .
\end{equation}

Now we find the symbol matrices $\Lambda(\hat{\mathbf J}_{13})$ for the vector of operators $\hat{\mathbf J}_{13}$, where 

\begin{equation}
[\hat{\Lambda}(\hat{\mathbf J}_{13})]_{\mu \nu} = \hat{U}^\dagger_{\alpha \mu} ( \hat{\mathbf J}_1 \delta_{\alpha \beta}) \hat{U}_{\beta \nu} + \hbar \, \hat{U}^\dagger_{\alpha \mu} {\mathbf S}_{\alpha \beta} \hat{U}_{\beta \nu} \, . 
\end{equation}
After converting the above to Weyl symbols, we find 

\begin{eqnarray*}
[\Lambda(\hat{\mathbf J}_{13})]_{\mu\nu}
&=& {\mathbf J}_1 \delta_{\mu \nu} - i \hbar U_{\alpha \mu}^* \{ U_{\alpha\mu } , \, {\mathbf J}_1 \}  + \hbar \, U_{\alpha \mu}^* {\mathbf S}_{\alpha \beta}  U_{\beta \nu} \\
&=& {\mathbf J}_1 \delta_{\mu \nu} - i \hbar \tau^{(\mu)*}_\alpha  \{ \tau^{(\nu)}_\alpha , \, {\mathbf J}_1 \}  + \hbar \, \tau^{(\mu)*}_\alpha {\mathbf S}_{\alpha \beta}  \tau^{(\nu)}_\beta \, . 
\end{eqnarray*}
See (4.10) in \cite{littlejohn1992} for a similar calculation. Let us denote the $i$th component of the second term above by $T^i_{\mu \nu}$, and use  Eq.\  (\ref{SLQN9j: eq_ tau_derivative}) and the orthogonality of $\tau$, 
\begin{equation}
\tau^{(\mu)*}_\alpha  \, \tau^{(\nu)}_\alpha = \delta_{\mu \nu} \, ,
\end{equation}
to get

\begin{eqnarray*}
T^i_{\mu \nu} &=&  - i \hbar \tau^{(\mu)*}_\alpha \{ \tau^{(\nu)}_\alpha  , \, J_{1i} \}  \\
&=&  - i  \hbar \tau^{(\mu)*}_\alpha \epsilon_{kji} J_{1k} \frac{\partial \tau^{(\nu)}_\alpha}{ \partial J_{1j}}   \\
&=& \hbar ({\mathbf A}_1^{(\mu)}  \times {\mathbf J}_1)_i \, \delta_{\mu \nu} + \hbar \frac{\mu J_{1i}}{I_1} \delta_{\mu \nu} - \hbar \, \tau^{(\mu)*}_\alpha S_{\alpha \beta}^i  \tau^{(\nu)}_\beta \, ,
\end{eqnarray*}
where in the second equality, we have used the reduced Lie-Poisson bracket ( (30) in \cite{littlejohn2007}) to evaluate the Poisson bracket $\{ \tau, {\mathbf J}_1 \}$, and where in the third equality, we have used  Eq.\  (\ref{SLQN9j: eq_ tau_derivative}) for $\partial \tau / \partial {\mathbf J}_1$. Notice the term involving ${\mathbf S}$ in $T^i_{\mu\nu}$ cancels out the same term in $\Lambda(\hat{\mathbf J}_{13})$, leaving us with a diagonal symbol matrix,

\begin{equation}
\label{SLQN9j: eq_J13_symbols}
[\Lambda(\hat{\mathbf J}_{13})]_{\mu\nu} = {\mathbf J}_1 \left[ 1 + \frac{\mu \hbar}{I_1} \right] + \hbar \, {\mathbf A}_1^{(\mu)} \times {\mathbf J}_1 \, . 
\end{equation}
See (4.12) in \cite{littlejohn1992}. Taking the square, we obtain 

\begin{equation}
[ \Lambda(\hat{\mathbf J}_{13}^2) ]_{\mu\nu}
=  ( I_1  + \mu \hbar)^2  \delta_{\mu \nu} \, ,
\end{equation}
Thus we see that $\Lambda(\hat{\mathbf J}_{13}^2)$ is diagonal. 

Finally, let us look at the last three remaining operators $\hat{\mathbf J}_{\text{tot}}$ in  Eq.\  (\ref{SLQN9j: eq_total_J_vector}). Since the symbols ${\mathbf J}_r$ for $r = 2,4,5$ Poisson commutes with ${\mathbf J}_1$, we find $\Lambda(\hat{\mathbf J}_r) = {\mathbf J}_r$. Using $\Lambda(\hat{\mathbf J}_{13})$ from  Eq.\  (\ref{SLQN9j: eq_J13_symbols}), we obtain 
 
\begin{eqnarray}
&& [\Lambda(\hat{\mathbf J}_{\text{tot}})]_{\mu\nu}   \\  \nonumber
&=& \left[ {\mathbf J}_1 \left( 1 + \frac{\mu \hbar}{I_1} \right) + \hbar \, {\mathbf A}_1^{(\mu)} \times {\mathbf J}_1 + ({\mathbf J}_2 + {\mathbf J}_4 + {\mathbf J}_5 ) \right]  \delta_{\mu \nu} \, .
\end{eqnarray}
Therefore all $\Lambda_i$, $i = 1,\dots, 10$ are diagonal. 

For each polarization $\mu$, not counting the eigenvalue equation for $S^2$, we have $9$ Hamilton-Jacobi equations associated with the $\Lambda_i$ in the $16$ dimensional phase space  ${\mathbb C}^8$. It turns out that not all of them are functionally independent. In particular, the Hamilton-Jacobi equations $\Lambda(\hat{I}_1) = I_1 - 1/2 \hbar = j_1 \hbar$ and $\Lambda(\hat{J}_{13}^2) = (I_1 + \mu \hbar)^2 = ( j_{13} + 1/2 )^2 \hbar^2$ are functionally dependent. The two equations are consistent only when we choose the polarization $\mu = j_{13} - j_1$. This reduces the number of independent Hamilton-Jacobi equations for $S(x)$ from $9$ to $8$, half of the dimension of the phase space ${\mathbb C}^8$. These $8$ equations define the Lagrangian manifold associated with the action $S(x)$ seen in Eq.\  (\ref{SLQN9j: eq_general_wavefunction}). 

Now let us restore the index $a$. We express the multicomponent wave-function $\psi^a_\alpha(x)$ in the form of  Eq.\  (\ref{SLQN9j: eq_general_wavefunction}), 

\begin{equation}
\label{SLQN9j: eq_general_wavefunction_a}
\psi^a_\alpha (x) = B_a(x) \, e^{i S_a(x) / \hbar} \,  \tau^a_\alpha(x, p) 
\end{equation}
Here the action $S_a(x)$ is the solution to the eight Hamilton-Jacobi equations associated with the $\mu^{\text{th}}$ entries $\lambda_i^a$ of $8$ of the symbol matrices $\Lambda_i^a$. 
\begin{eqnarray}
\label{SLQN9j: eq_Hamilton_Jacobi_a}
I_1 &=& (j_1 + 1/2) \hbar \, ,  \\    \nonumber
I_2 &=& (j_2 + 1/2) \hbar \, ,  \\  \nonumber
I_4 &=& (j_4 + 1/2) \hbar \, ,  \\  \nonumber
I_5 &=& (j_5 + 1/2) \hbar \, ,  \\  \nonumber
J_{24}^2 &=& (j_{24} + 1/2)^2 \hbar^2 \, ,  \\  \nonumber
{\mathbf J}_{\text{tot}}^{(a)} &=& {\mathbf J}_1 \left[ 1 + \frac{\mu \hbar}{I_1} \right] + \hbar \, {\mathbf A}_1 \times {\mathbf J}_1 + ({\mathbf J}_2 + {\mathbf J}_4 + {\mathbf J}_5 ) = {\mathbf 0} \, ,
\end{eqnarray}
and $\tau^a = \tau^{(\mu)}$ with $\mu = j_{13}-j_1$. Note that both $\tau^a$ and $S_a$ involve the vector potential ${\mathbf A}_1$, and so are gauge dependent quantities.

We carry out an analogous analysis for $\psi^b(x)$. The result is
\begin{equation}
\label{SLQN9j: eq_general_wavefunction_b}
\psi^b_\alpha (x) = B_b(x) \, e^{i S_b(x) / \hbar} \,  \tau^b_\alpha(x, p) \, ,
\end{equation}
where $S_b(x)$ is the solution to the following $8$ Hamilton-Jacobi equations, 

\begin{eqnarray}
I_1 &=& (j_1 + 1/2) \hbar \, ,  \\    \nonumber
I_2 &=& (j_2 + 1/2) \hbar \, ,  \\  \nonumber
I_4 &=& (j_4 + 1/2) \hbar \, ,  \\  \nonumber
I_5 &=& (j_5 + 1/2) \hbar \, ,  \\  \nonumber
J_{12}^2 &=& (j_{12} + 1/2)^2 \hbar^2 \, ,  \\  \nonumber
{\mathbf J}_{\text{tot}}^{(b)} &=& {\mathbf J}_4 \left[ 1 + \frac{\nu \hbar}{I_4} \right] + \hbar \, {\mathbf A}_4 \times {\mathbf J}_4 + ({\mathbf J}_1 + {\mathbf J}_2 + {\mathbf J}_5 ) = {\mathbf 0} \, ,
\end{eqnarray}
and the spinor $\tau^b = \tau_b^{(\nu)}$,  $\nu = j_{34}-j_4$, satisfies 

\begin{equation}
\label{SLQN9j: eq_JS_eigenvector_b}
({\mathbf J}_4 \cdot {\mathbf S})_{\alpha \beta} \, ( \tau^{(\nu)}_b)_\beta = \nu I_4 \, ( \tau^{(\nu)}_b)_\beta \, .
\end{equation}
The vector potential ${\mathbf A}_4$ is defined by

\begin{equation}
{\mathbf A}_4 = i ( \tau^b )^\dagger \, \frac{\partial \tau^b }{\partial {\mathbf J}_4} \, .
\end{equation}
Again,  both $\tau^b$ and $S_b$ involve the vector potential ${\mathbf A}_4$, and so are gauge dependent quantities.

\section{\label{SLQN9j: sec_factorization} A GAUGE-INVARIANT FORM OF THE MULTICOMPONENT WAVE-FUNCTIONS}

In this section, we derive a gauge-invariant factorization for $\psi^a(x)$. In general, an action $S(x)$ is defined by the integral of $p dx$ on a Lagrangian manifold. The action $S_a(x)$ is associated with the Lagrangian manifold ${\cal L}_a$, the level set defined by  Eq.\  (\ref{SLQN9j: eq_Hamilton_Jacobi_a}). We will express $S_a(x)$ in terms of an action on a nearby gauge-invariant Lagrangian manifold. First, we pick a reference point $z_0$ on ${\mathcal L}_a$ and let $x_0$ be the $x$ coordinates of $z_0$, and write $S_a(x) = \int_\gamma p dx$ as a line integral along a path $\gamma$ that connects $x_0$ to $x$. Because ${\mathcal L}_a$ is Lagrangian, $S_a(x)$ is independent of the path $\gamma$ chosen. Then we define an intermediate nearby gauge-invariant manifold ${\cal L}_a'$ using a set of gauge invariant coordinates $z'$, which we will define in the next paragraph. We then express the line integral $S_a(x)$ on ${\cal L}_a$ as a perturbation of a line integral $S_a'(x) = \int_\gamma p'(x) dx$, where $p'(x)$ is the coordinate representation of ${\cal L}_a'$. Because ${\cal L}_a'$ is not Lagrangian, $S_a'(x)$ will depend on the path $\gamma$. Finally, we express the line integral $S_a'(x)$ on ${\cal L}_a'$ as a perturbation of a known action $S_a^{6j}(x)$, which is supported on another nearby gauge-invariant Lagrangian manifold ${\cal L}^{6j}_a$, the so called A-manifold  from \cite{littlejohn2010b}. This manifold appears in the analysis of the Ponzano-Regge action of the $6j$-symbol.  We then show that the perturbations are independent of the path $\gamma$ chosen, by combining the extra phases generated from these perturbations with the spinor field, and show that the result becomes a field of quantum rotations of a fixed reference spinor on the Lagrangian manifold $S_a^{6j}(x)$. We show the advantages of our factorization by deriving a simple expression for the inner products between two multicomponent wave-functions. 

\subsection{The intermediate gauge-invariant manifold ${\cal L}'$ }

The gauge invariant coordinates $z'$ were introduced in \cite{littlejohn1991, littlejohn1992} to formulate gauge invariant quantization conditions, and to show the overall wave function $\psi(x)$ in  Eq.\  (\ref{SLQN9j: eq_general_wavefunction}) is invariant under a local $U(1)$ gauge transformation. We use the coordinate map from $z$ to $z'$ here to define a gauge invariant intermediate manifold ${\cal L}_a'$. Note the notation $z'$ in \cite{littlejohn1992} denotes vectors of real $(x, p)$ coordinates, here $z'$ refers to complex coordinates. We define the near-identity map $Z$ as follows, 

\begin{equation}
\label{SLQN9j: eq_z_prime}
z_{r\mu}' = Z_{r\mu}(z_{r\mu}) = z_{r\mu} - i \hbar (\tau^a)^\dagger \left(\frac{\partial \tau^a}{\partial z_{r\mu}}  \right) \, ,
\end{equation}
where $Z_{r\mu}$ are the functions, and $z_{r\mu}'$ are the values of the functions evaluated at $z_{r\mu}$. Since $\tau^a$ only depends on ${\mathbf J}_1$, the derivatives $\partial \tau^a / \partial z_{r\mu}$ for $r = 2,4,5$ vanish, and we have

\begin{equation}
z_{r\mu}' = Z_{r\mu} (z_{r\mu}) = z_{r\mu} \, ,   \quad \quad \quad   r = 2,4,5.
\end{equation}
For $r = 1$, we use the chain rule,  Eq.\  (\ref{SLQN9j: eq_I_Ji_symbols}), and  Eq.\  (\ref{SLQN9j: eq_gauge_pot_A}) to find

\begin{eqnarray*}
z_{1\mu}' = Z_{1\mu} (z_{r\mu}) &=& z_{1\mu} - i \hbar \tau^\dagger \left(\frac{\partial \tau}{\partial {\mathbf J}_1} \right) \cdot \frac{\partial {\mathbf J}_1}{z_{1\mu}}  \\
&=& z_{1\mu} - i \hbar {\mathbf A}_1 \cdot \left( \frac{1}{2} \, \sum_{\nu} \, {\bm \sigma}_{\mu\nu} \, z_{1\nu} \right) \, .
\end{eqnarray*}
Using the identity 

\begin{equation}
\sigma_i \, \sigma_j = \delta_{ij} + i \epsilon_{ijk} \sigma_k \, ,
\end{equation}
we obtain to first order 

\begin{eqnarray}
J_{1j} \circ Z
&=& \frac{1}{2}  \, \overline{z}'  \sigma_i  z' \, ,   \nonumber  \\
&=& \left[  \overline{z}_{1} +  \frac{i \hbar}{2} { A}_{1i}   \overline{z}_1 \, \sigma_i  \right] \sigma_j  \left[  z_{1} -  \frac{i \hbar}{2} { A}_{1k}  \sigma_k \, z_{1} \right]  \nonumber  \\
&=& \frac{1}{2}  \overline{z}_1 \sigma_j  z_{1} - \frac{1}{2} \hbar \epsilon_{ijk} { A}_{1i} { J}_{1k}  + \frac{1}{2} \hbar \epsilon_{ijk} { A}_{1k} { J}_{1i}  \nonumber  \\
&=& { J}_{1j} + \hbar ( {\mathbf A}_1 \times {\mathbf J}_1 )_j \, .
\end{eqnarray}
To follow these manipulations, one must pay close attention to the difference between the functions and the values of the functions. 

Looking at the last Hamilton-Jacobi equation from  Eq.\  (\ref{SLQN9j: eq_Hamilton_Jacobi_a}), we see that the gauge dependence of ${\mathbf J}_{\text{tot}}$ is captured by the coordinate transformation $Z$, that is, 

\begin{equation}
\label{SLQN9j: eq_Hamilton_Jacobi_Jtot_prime}
{\mathbf J}_{\text{tot}}^{(a)} = \left[ 1 + \frac{\mu \hbar}{I_1} \right]  {\mathbf J}_1 \circ Z  + ({\mathbf J}_2 + {\mathbf J}_4 + {\mathbf J}_5 ) = {\mathbf 0} \, .
\end{equation}
Therefore, it is tempting define a new gauge-invariant manifold by dropping $Z$ from  Eq.\  (\ref{SLQN9j: eq_Hamilton_Jacobi_Jtot_prime}). This is equivalent to treating $Z$ as an active map that maps points on ${\cal L}_a$ to a new gauge-invariant manifold ${\cal L}_a'$ that satisfies the following Hamilton-Jacobi equations, 

\begin{eqnarray}
\label{SLQN9j: eq_Hamilton_Jacobi_a_prime}
I_1 &=& (j_1 + 1/2) \hbar \, ,  \\    \nonumber
I_2 &=& (j_2 + 1/2) \hbar \, ,  \\  \nonumber
I_4 &=& (j_4 + 1/2) \hbar \, ,  \\  \nonumber
I_5 &=& (j_5 + 1/2) \hbar \, ,  \\  \nonumber
J_{24}^2 &=& (j_{24} + 1/2)^2 \hbar^2 \, ,  \\  \nonumber
{\mathbf J}_{\text{tot}}^{(a)'} &=& {\mathbf J}_1 \left[ 1 + \frac{\mu \hbar}{I_1} \right]  + ({\mathbf J}_2 + {\mathbf J}_4 + {\mathbf J}_5 ) = {\mathbf 0} \, .
\end{eqnarray}
It is easy to check that $I_1 \circ Z = I_1$ to first order, and the remaining four equations are not affected by $Z$. 

In other words, if $(x, p(x)) \in {\mathcal L}_a$, then $(x', p') = Z(x, p(x)) \in {\mathcal L}_a'$ and vice versa. Here $(x, p(x))$ is the coordinate representation of the Lagrangian manifold ${\mathcal L}_a$.  We let $(x, p'(x))$ denote the coordinate representation of ${\mathcal L}_a'$.

\subsection{First perturbation of the action}

Now let us relate the line integral $S_a(x)$ to the line integral $S_a'(x) = \int_\gamma p'(x) dx$. We express the map $Z$ from Eq.\ (\ref{SLQN9j: eq_z_prime}) in terms of the $x, p$ coordinates, 
\begin{eqnarray}
\label{prime_coord_p}
p_{1\mu}'  &=&  p_{1\mu} -  i \hbar (\tau^a)^\dagger \, \frac{\partial \tau^a}{\partial x_{1\mu}} \, , \\
\label{prime_coord_x}
x_{1\mu}'  &=&  x_{1\mu} +  i \hbar (\tau^a)^\dagger \, \frac{\partial \tau^a}{\partial p_{1\mu}} \, .
\end{eqnarray}
To find the value of $p'(x)$ in terms of $x$, $p(x)$, use the relations $p' = p'(x')$ and $(x', p'(x')) = Z(x, p(x))$, as follows,   

\begin{equation}
p (x) - i \hbar (\tau^a)^\dagger \, \frac{\partial \tau^a}{\partial {x} }  = p' = p'(x') = p' \left( x + i \hbar  (\tau^a)^\dagger \, \frac{\partial \tau^a}{\partial p}   \right) \, . 
\end{equation}
Expanding in powers of $\hbar$, we find

\begin{eqnarray}
p_{1\mu}'(x) &=& p_{1\mu}(x) - i \hbar ( \tau^a )^\dagger \,  \left[ \frac{\partial \tau^a}{\partial x_{1\mu}} + \frac{\partial \tau^a}{\partial p_{1\nu} } \; \frac{\partial p_{1\nu}}{\partial x_{1\mu}} \right]   \nonumber    \\
&=& p_{1\mu}(x) - i \hbar ( \tau^a )^\dagger \, \frac{d \tau^a}{d x_{1\mu}} \, , 
\end{eqnarray}
and $p_{r \mu}'(x) = p_{r \mu}(x)$ for $r = 2,4,5$. Integrating along the path $\gamma$, we obtain the relation

\begin{eqnarray}
S_a(x) &=& S_a'(x) + i \hbar \int_{\gamma} \, ( \tau^a )^\dagger \, \frac{d \tau^a}{d {\mathbf x}_1} d {\mathbf x}_1 \nonumber  \\
\label{SLQN9j: eq_perturbation1}
&=&  S_a'(x) + \hbar \int_{\gamma} \, {\mathbf A}_1 \cdot d{\mathbf J}_1 \, ,
\end{eqnarray}
where we have used the definition of ${\mathbf A}_1$ in  Eq.\  (\ref{SLQN9j: eq_gauge_pot_A}) in the second equality.

\subsection{Second perturbation of the action}

We now express the line integral $S'(x)$ as a perturbation of a known action $S_a^{6j}(x)$ on a nearby gauge-invariant Lagrangian manifold ${\mathcal L}_a^{6j}$. The Lagrangian manifold ${\cal L}_a^{6j}$ is defined by the following Hamilton-Jacobi equations: 

\begin{eqnarray}
\label{SLQN9j: eq_Hamilton_Jacobi_6j}
I_1 &=& (j_1 + 1/2) \hbar \, ,    \nonumber \\    \nonumber
I_2 &=& (j_2 + 1/2) \hbar \, ,  \\  \nonumber
I_4 &=& (j_4 + 1/2) \hbar \, ,  \\  \nonumber
I_5 &=& (j_5 + 1/2) \hbar \, ,  \\  \nonumber
J_{24}^2 &=& (j_{24} + 1/2)^2 \hbar^2 \, , \\ 
{\mathbf J}_{\text{tot}} &=& {\mathbf J}_1  + {\mathbf J}_2 + {\mathbf J}_4 + {\mathbf J}_5 = {\mathbf 0}  \, .
\end{eqnarray}

Now we find the difference between $S_a'(x)$ and $S_a^{6j}(x)$. By lifting $\gamma$ to two nearby paths $\gamma_1$ and $\gamma_1'$ on ${\cal L}_a^{6j}$ and  ${\cal L}_a'$, respectively, we can model the two nearby paths $\gamma_1$ and $\gamma_1'$ by Hamiltonian flows generated by the Hamiltonians in Eq.\  (\ref{SLQN9j: eq_Hamilton_Jacobi_6j}) and Eq.\  (\ref{SLQN9j: eq_Hamilton_Jacobi_a_prime}), respectively. Since the first five Hamiltonians $(I_1, I_2, I_4, I_5, J_{24}^2)$ are identical in Eq.\  (\ref{SLQN9j: eq_Hamilton_Jacobi_6j}) and Eq.\  (\ref{SLQN9j: eq_Hamilton_Jacobi_a_prime}), we will focus on the segments of $\gamma_1$ and $\gamma_1'$ that are  generated by $\hat{\mathbf n} \cdot {\mathbf J}_{\text{tot}}$ and $\hat{\mathbf n} \cdot {\mathbf J}_{\text{tot}}^{(a)'}$. Here $\hat{\mathbf n}$ is some fixed unit vector, and $\theta$ is the parameter along the flows of the two segments. Let us denote the projection of the two segments onto the $x$ space by $x(\theta)$ and $x'(\theta)$, respectively. We assume $S_a'(x'(\theta)) = S_a'(x(\theta))$ holds to first order.

The equations of motion for the Hamiltonian $\hat{\mathbf n} \cdot {\mathbf J}_{\text{tot}}$ are

\begin{equation}
\label{SLQN9j: eq_ EOM_Jtot}
\frac{d  \overline{z}_{r\mu}}{d\theta} =  \frac{\partial (\hat{\mathbf n} \cdot {\mathbf J}_{\text{tot}})}{ \partial ( - i z_{r\mu})} =  \frac{i}{2} \, \sum_\nu  \overline{z}_{r\nu} (\hat{\mathbf n} \cdot {\bm \sigma})_{\nu\mu} \, ,
\end{equation}
so the line integral is
 
\begin{equation}
\label{SLQN9j: eq_dS_6j}
S^{6j}(\theta) = \text{Im} \, \int_0^\theta \sum_{r\nu}  z_{r\nu} d\overline{z}_{r\nu} = \int_0^\theta (\hat{\mathbf n} \cdot {\mathbf J}_{\text{tot}}) \, d\theta = 0 \, ,
\end{equation}
where we have used  Eq.\  (\ref{SLQN9j: eq_ EOM_Jtot}) for $d\overline{z}$ and the definitions for ${\mathbf J}_r$ in  Eq.\  (\ref{SLQN9j: eq_I_Ji_symbols}) in the second equality, and where we have used ${\mathbf J}_{\text{tot}} = {\mathbf 0}$ on ${\cal L}_a^{6j}$ in the last equality. Thus the action generated by part of the flow in $\gamma_1$ vanishes. 

The equations of motion for the Hamiltonian $\hat{n} \cdot {\mathbf J}_{\text{tot}}^{(a)'}$ are the following, 

\begin{equation}
\label{SLQN9j: eq_ EOM_Jtot_prime}
\frac{d  \overline{z}_{r\nu}}{d\theta} =  \frac{i}{2} \sum_\rho \,  \overline{z}_{r\rho} (\hat{\mathbf n} \cdot {\bm \sigma})_{\rho\nu} \left(1 + \frac{\mu \hbar}{I_1} \right) -  \frac{i \mu \hbar }{2 I_1^2}(\hat{\mathbf n} \cdot \hat{\mathbf J}_1 ) \overline{z}_{1\nu} \, , 
\end{equation}
so the line integral is

\begin{eqnarray}
\label{SLQN9j: eq_dS_prime}
S_a'(\theta) 
&=& \int_0^\theta (\hat{\mathbf n} \cdot  {\mathbf J}_{\text{tot}}^{(a)'} ) \, d\theta  - \frac{\mu \hbar}{I_1} \int_0^\theta (\hat{\mathbf n} \cdot {\mathbf J}_1) \, d\theta   \nonumber  \\
&=& - \frac{\mu \hbar}{I_1} \int_0^\theta (\hat{\mathbf n} \cdot {\mathbf J}_1) \, d\theta  \, ,
\end{eqnarray}
where we have used  Eq.\  (\ref{SLQN9j: eq_ EOM_Jtot_prime}) and Eq.\  (\ref{SLQN9j: eq_I_Ji_symbols}) in the first equality, and the last equation of  Eq.\  (\ref{SLQN9j: eq_Hamilton_Jacobi_a_prime}) in the second equality. Thus the part of line integral along $\gamma_1'$ does not vanish. 

From the difference of the line integrals in  Eq.\  (\ref{SLQN9j: eq_dS_6j}) and Eq.\  (\ref{SLQN9j: eq_dS_prime}), we obtain 

\begin{equation}
\label{SLQN9j: eq_perturbation2}
S_a'(x) = S_a^{6j}(x) - \sum_i \frac{\mu \hbar}{I_1}   \int_{\gamma_1^i}  \hat{\mathbf n}_i \cdot {\mathbf J}_1 \, d\theta \, .
\end{equation}
where the sum in the last term is over the segments in $\gamma_1^i$ of $\gamma_1$ that are generated by Hamiltonians of the type $\hat{\mathbf n}_i \cdot {\mathbf J}_{\text{tot}}$.

Combining the two perturbations  Eq.\  (\ref{SLQN9j: eq_perturbation1}) and Eq.\  (\ref{SLQN9j: eq_perturbation2}), we can express the action $S_a(x)$ on the gauge-dependent Lagrangian manifold ${\mathcal L}_a$ as a perturbation of a known action $S_a^{6j}(x)$ on the gauge-invariant Lagrangian manifold ${\mathcal L}_a^{6j}$, 

\begin{equation}
\label{SLQN9j: eq_perturbation}
S_a(x) = S_a^{6j}(x) + \hbar \int_{\gamma} \, {\mathbf A}_1 \cdot d{\mathbf J}_1 -  \sum_i  \frac{\mu \hbar}{I_1}   \int_{\gamma_1^i}  \hat{\mathbf n}_i \cdot {\mathbf J}_1 \, d\theta \, .
\end{equation}

We note that the second perturbation in  Eq.\  (\ref{SLQN9j: eq_perturbation}) vanishes unless $\gamma_1$ contains flows generated by the total angular momentum. Moreover, since the only Hamiltonian in  Eq.\  (\ref{SLQN9j: eq_Hamilton_Jacobi_6j}) that geenerate flows that change the values of ${\mathbf J}_1$ or ${\mathbf J}_5$ is the total angular momentum, the first perturbation in  Eq.\  (\ref{SLQN9j: eq_perturbation}) also vanishes unless $\gamma$ contains flows generated by the total angular momentum. In other words, only the flows of the Hamiltonians ${\mathbf n} \cdot {\mathbf J}_{\text{tot}}$ generate the two perturbations in Eq.\  (\ref{SLQN9j: eq_perturbation}).

Since $S_a(x)$ and $S_a^{6j}(x)$ are actions on Lagrangian manifolds, their values are independent of the path $\gamma$ we choose to perform this perturbation. Thus, the last two terms in Eq.\  (\ref{SLQN9j: eq_perturbation}) together should also be independent of the path $\gamma_1$. We show this is indeed the case by rewriting the spinor field on the Lagrangian manifold ${\mathcal L}_a^{6j}$, which is independent of $\gamma_1$, in terms of these two terms and a field of rotational matrices on ${\mathcal L}_a^{6j}$, which is also independent of $\gamma_1$, and a fixed spinor.

\subsection{Transformation of the spinor field}

We now turn our attention to the spinor $\tau^a(z)$ in the multicomponent wave-function from  Eq.\  (\ref{SLQN9j: eq_general_wavefunction}). Let $z_0$ be the reference point and let $\gamma_1$ be the path from $z_0$ to $z$ on ${\cal L}_a^{6j}$, as defined above. If $\gamma_1$ is generated by  the Hamiltonian flows of $(I_1, I_2, I_4, I_5,  J_{24}^2)$, which do not change the value of ${\mathbf J}_1$, we have $d{\mathbf J}_1 = 0$ along $\gamma_1$. Since the spinor $\tau^a(z)$ only depend on ${\mathbf J}_1$, we find

\begin{equation}
\tau^a(z) = \tau^a(z_0) \, . 
\end{equation}
If $\gamma_1$ is generated by the Hamiltonian flows of ${\mathbf n} \cdot {\mathbf J}_{\text{tot}}$ by an angle $\theta$, then $d {\mathbf J}_1$ is nonzero, and is given by

\begin{equation}
\label{SLQN9j: eq_dJ1}
d{\mathbf J}_1 = {\mathbf n} \times {\mathbf J}_1 d\theta \, .
\end{equation}
Now we analyze how a spinor section $\tau^a(z)$, such as the one from  Eq.\  (\ref{SLQN9j: eq_tau_NSG}), changes as we follow $\gamma_1$. To express the gauge dependence of $\tau^a(z)$ on the gauge potential ${\mathbf A}_1$, we find it convenient to express $\tau^a(z)$ in terms of the parallel transport of the spinor along $\gamma_1$, that is, 

\begin{equation}
\label{SLQN9j: eq_SLQN9j: eq_tau_factorization_1}
\tau^a(z) = \text{exp} \left\{-i \int_{\gamma_1} \, {\mathbf A}_1 \cdot d {\mathbf J}_1 \right\} \, \tilde{\tau}^a(z) \, . 
\end{equation}
Here $\tilde{\tau}^a$ is the solution to the parallel transport equation along $\gamma$,  

\begin{equation}
\label{SLQN9j: eq_parallel_transport}
\frac{d \tilde{\tau}^a}{d\theta} = \frac{i}{{\mathbf J}_1^2} \, \left[ \left(\frac{d {\mathbf J}_1}{d\theta} \times {\mathbf J}_1 \right) \cdot {\mathbf S} \right] \tilde{\tau}^a(\theta)
\end{equation}
with the initial condition $\tilde{\tau}^a(0) = \tau^a(z_0)$. We see that the phase in  Eq.\  (\ref{SLQN9j: eq_SLQN9j: eq_tau_factorization_1}) cancels the first term in  Eq.\  (\ref{SLQN9j: eq_perturbation}). 

To generate a phase that will cancel the second perturbation in  Eq.\  (\ref{SLQN9j: eq_perturbation}), we rewrite the parallel transported spinor $\tilde{\tau}^a$ in terms of quantum rotation matrices. Let us define 
\begin{equation}
U(\hat{\mathbf n}, d\theta) = e^{- i (\hat{\mathbf n} \cdot {\mathbf S}) d\theta} \, ,
\end{equation}
which has the same axis and angle as those in $\gamma$. We decompose the axis $\hat{\mathbf n}$ into components along and perpendicular to ${\mathbf J}_1$, 

\begin{equation}
\hat{\mathbf n} = \frac{\hat{\mathbf n} \cdot {\mathbf J}_1}{J_1^2} \, {\mathbf J}_1 + \frac{{\mathbf J}_1 \times (\hat{\mathbf n} \times {\mathbf J}_1)}{J_1^2} \, . 
\end{equation}
Using the fact that $\tilde{\tau}^a({\mathbf J}_1)$ is an eigenvector of ${\mathbf J}_1 \cdot {\mathbf S}$ from  Eq.\  (\ref{SLQN9j: eq_JS_eigenvector}), and the parallel transport equation from  Eq.\  (\ref{SLQN9j: eq_parallel_transport}), with $d{\mathbf J}_1$ defined in  Eq.\  (\ref{SLQN9j: eq_dJ1}), we find

\begin{eqnarray}
&& U(\hat{\mathbf n}, d \theta) \, \tilde{\tau}^a(z) \nonumber  \\  \nonumber
&=& \text{exp} \left\{ -i \frac{{\mathbf J}_1 \times (\hat{\mathbf n} \times {\mathbf J}_1) }{J_1^2} \, \cdot {\mathbf S} \, d \theta \right\} \,  \\  \nonumber
&& \quad \quad \quad \text{exp} \left\{ -i \frac{\hat{\mathbf n} \cdot {\mathbf J}_1}{J_1^2} \, {\mathbf J}_1 \cdot {\mathbf S} \, d \theta \right\} \, \tilde{\tau}^a(z)  \\  \nonumber
&=&  e^{i \gamma} \, \left[ \text{exp} \left\{-i \frac{{\mathbf J}_1 \times \, d{\mathbf J}_1 }{J_1^2} \, \cdot {\mathbf S} \right\}  \tilde{\tau}^a(z) \, \right] \\
&=& e^{i \gamma}  \, \tilde{\tau}^a(z+ d z) \, ,
\end{eqnarray}
where 

\begin{equation}
e^{i \gamma}  =  \text{exp}  \left\{-i \mu \frac{\hat{\mathbf n} \cdot {\mathbf J}_1}{J_1} \, d \theta \right\} \, .
\end{equation}
Moving the exponential to the other side, and integrating with respect to $d\theta$ along $\gamma$ from $z_0$ to $z$, we find an expression for the parallel transport of the spinor in terms of the rotation matrix, as follows, 

\begin{equation}
\tilde{\tau}^a (z) = \text{exp} \left\{i \mu \int_\gamma  \frac{\hat{\mathbf n} \cdot {\mathbf J}_1}{J_1} \, d \theta \right\}  \, U(\hat{\mathbf n}, \, \theta) \, \tau^a(z_0) \, .
\end{equation}
In general, if $\gamma_1$ consists of segments generated by overall rotations, say $\gamma = \prod_i \gamma_1^i$, then by applying the above formula sequentially, we have 

\begin{equation}
\label{SLQN9j: eq_tau_tilde_factorization}
\tilde{\tau}^a (z) = \text{exp} \left\{i \mu \int_\gamma  \frac{\hat{\mathbf n}(\gamma) \cdot {\mathbf J}_1}{J_1} \, d \theta_\gamma \right\}  \, \prod_{\gamma_1^i} \, U(\hat{\mathbf n}_i, \, \theta_i) \, \tau^a(z_0)
\end{equation}
Putting Eq.\ (\ref{SLQN9j: eq_tau_tilde_factorization}) back into Eq.\ (\ref{SLQN9j: eq_SLQN9j: eq_tau_factorization_1}), we have

\begin{eqnarray}
 \tau^a (z)  
&=&  \text{exp} \left\{ -i  \int_\gamma \, {\mathbf A}_1 \cdot d{\mathbf J}_1 + i \frac{\mu}{J_1}    \int_\gamma  \frac{\hat{\mathbf n}(\gamma) \cdot {\mathbf J}_1}{J_1} \, d \theta_\gamma   \right\}   \nonumber  \\
\label{SLQN9j: eq_tau_factorization}
&& \quad \left[ \prod_{\gamma_1^i} U(\hat{\mathbf n}_i, \, \theta_i)  \right] \, \tau^a(z_0) \, . 
\end{eqnarray}
We note that the product of rotation matrices are almost uniquely defined by the end points $z_0$ and $z$. To see this, project this series of rotations onto a series of classical rotations $R_i$ in $SO(3)$. Because both ${\mathbf J}_1$ and ${\mathbf J}_5$ are invariant under the other Hamiltonians $(I_1, I_2, I_4, I_5, J_{24}^2)$, the triangle 1-5-24 can only change orientation under the flows of the total angular momentum. Thus, this series of rotations $R_i$ must take the triangle 1-5-24 at $z_0$ to the triangle 1-5-24 at $z$. There is a unique rotation $R$ that does this. Thus the product of the rotation matrices $U(\hat{\mathbf n}_i, \, \theta_i)$ is equal to a lift $U$ of the $SO(3)$ rotation matrix $R$. We can write

\begin{equation}
U_a(x) = \prod_{\gamma_1^i} U({\mathbf n}_i, \, \theta_i) \, .
\end{equation}

We conclude that both the spinor field $\tau^a(z)$ and the field of rotation matrices $U(z)$ are independent of the path $\gamma_1$. As a result, the phase factors  in Eq.\  (\ref{SLQN9j: eq_tau_factorization}), and the perturbations in Eq.\  (\ref{SLQN9j: eq_perturbation}) are independent of the path $\gamma_1$.

When we put Eq.\  (\ref{SLQN9j: eq_perturbation}) and Eq.\  (\ref{SLQN9j: eq_tau_factorization}) together into the multicomponent wave-function $\psi^a(x) = B(x) e^{i S_a(x) / \hbar} \, \tau^a(x) $ the phase factors exactly cancel out. In the end, we obtain a gauge-invariant representation of the wave-function,  

\begin{equation}
\label{SLQN9j: eq_wave_fctn_factorization_a}
\psi^a(x) = B_a(x) \, e^{i S_a^{6j}(x) / \hbar} \, U_a(x)  \, \tau^a(x_0) \, . 
\end{equation}

Similarly, the multicomponent wave-function for the state $\ket{b}$ has the following form,

\begin{equation}
\label{SLQN9j: eq_wave_fctn_factorization_b}
\psi^b(x) = B_b(x) \, e^{i S_b^{6j}(x) / \hbar} \,  U_b(x)   \, \tau^b(x_0) \, , 
\end{equation}
where for convenience, we have chosen the same reference point $z_0$. This is possible, because the reference points are arbitrary on both manifolds, and we can choose $z_0$ to lie in an intersection of ${\mathcal L}_a^{6j}$ and ${\mathcal L}_b^{6j}$.

\subsection{the inner products of multicomponent wave-functions}

We now take the inner product between the two multicomponent wave-functions $\psi^a(x)$ and $\psi^b(x)$, and perform a stationary phase approximation. From Eq.\ (\ref{SLQN9j: eq_wave_fctn_factorization_a}) and Eq.\  (\ref{SLQN9j: eq_wave_fctn_factorization_b}), we find

\begin{eqnarray}
\label{SLQN9j: eq_inner_product}
 \braket{b|a} 
&=& \int \, dx B_a(x) B_b(x) \, e^{i ( S_a^{6j}(x) - S_b^{6j}(x)) / \hbar}    \\  \nonumber
&& ( \tau^b(x_0) )^\dagger [U_b(x)]^\dagger \, U_a(x) \, \tau^a(x_0) \, . 
\end{eqnarray}
Let ${\cal M}_k$ denote the $k^{\text{th}}$ component of the intersection set of the two Lagrangian manifolds ${\cal L}_a^{6j}$ and ${\cal L}_b^{6j}$. Choose the reference point $z_0 \in {\cal M}_0$. We know the difference of the two actions $S_b(x_k) - S_a(x_k)$ is constant in each component ${\cal M}_k$, since the integrals of $p dx$ cancel in the intersections. In particular, since $S_b(x_0) = S_a(x_0) = 0$, it vanishes in ${\cal M}_0$. Similarly, by their unitary property, the product of the rotation matrices cancel out in the intersections, and is constant in each ${\cal M}_k$. In particular, it is the identity matrix in ${\cal M}_0$.

Suppose we know the result of the stationary phase approximation to the integral in  Eq.\  (\ref{SLQN9j: eq_inner_product}), when the spinors and rotation matrices are absent, is given by 

\begin{eqnarray}
\label{SLQN9j: eq_inner_product_wo_spinor}
&& \int \, dx B_a(x) B_b(x) \, e^{i ( S_a^{6j}(x) - S_b^{6j}(x)) / \hbar}   \\   \nonumber
&& = e^{i\kappa} \sum_k    \Omega_k  \, \text{exp} \{i [S_a^{6j}(x_k) - S_b^{6j}(x_k) - \mu_k \pi /2 ] / \hbar \} \, , 
\end{eqnarray}
then by treating the spinors as slowly varying and evaluating them at the stationary points, we get

\begin{eqnarray}
 \braket{b|a}  
&=&  e^{i\kappa} \sum_k    \Omega_k  \, \text{exp} \{i [S_a^{6j}(z_k) - S_b^{6j}(z_k) - \mu_k \pi /2 ] / \hbar \}  \nonumber  \\
\label{SLQN9j: eq_general_formula}
&& \left(U_b^{0k} \tau^b(z_0)\right)^\dagger  \left(U_a^{0k} \tau^a(z_0)\right) \,  .
\end{eqnarray}
In the above formula, the sum is over the components of the intersection set, and $z_k$ is a point in the $k$th component, and where $U_a^{0k}$ is a rotation matrix determined from the rotation that takes $z_0$ to a point $z_k \in {\cal M}_k$ inside ${\cal L}_a^{6j}$, and $U_b^{0k}$ is similarly defined. The formula  Eq.\  (\ref{SLQN9j: eq_general_formula}) is independent of the choice of $z_k$, because any other choice $z_k'$ will multiply both $U_a^{0j}$ and $U_b^{0j}$ by the same additional rotation matrix generated from a path from $z_k$ to $z_k'$, which cancel out in the product $(U_b^{0k})^\dagger U_a^{0k}$. We see that our factorization of the multicomponent wave-function leads to expression,  Eq.\  (\ref{SLQN9j: eq_general_formula}), for the inner product between two multicomponent wave-functions.

\section{\label{SLQN9j: sec_9j_derivation}  DERIVATION FOR THE $9J$-SYMBOL FORMULA}

We derive the main result  Eq.\  (\ref{SLQN9j: eq_main_formula}) for the $9j$-symbol by using the general formula  Eq.\  (\ref{SLQN9j: eq_general_formula}). The stationary phase approximation without the spinors in  Eq.\  (\ref{SLQN9j: eq_inner_product_wo_spinor}) was calculated in an earlier paper on the $6j$-symbol \cite{littlejohn2010}. In the following, we summarize the results of that calculation, and then evaluate the spinor inner products at the stationary phase points to derive the inner product of the two multicomponent wave-functions, as in  Eq.\  (\ref{SLQN9j: eq_general_formula}). From this inner product, we derive the asymptotic formula for the $9j$-symbol. 

\subsection{The relative action at the two stationary sets}

\begin{figure}[tbhp]
\begin{center}
\includegraphics[width=0.45\textwidth]{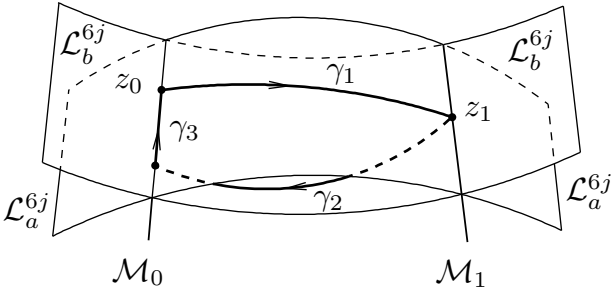}
\caption{Illustration for the intersection of the Lagrangian manifolds, and for the paths $\gamma_1$, $\gamma_2$, and $\gamma_3$.}
\label{SLQN9j: fig_lagrangian_manifold}
\end{center}
\end{figure}

Let us describe the Lagrangian manifolds and the relevant paths used to calculate their actions. The two Lagrangian manifolds ${\cal L}_a^{6j}$ and ${\cal L}_b^{6j}$ intersect at two sub-manifolds ${\cal M}_0$ and ${\cal M}_1$. The relative action $S_a^{6j}(z_1) - S_b^{6j}(z_1)$ is calculated along a loop $\gamma = \gamma_1 + \gamma_2 + \gamma_3$, as illustrated in Fig.\ \ref{SLQN9j: fig_lagrangian_manifold}. First it goes from a point $z_0$ in ${\cal M}_0$ to a point $z_1$ in ${\cal M}_1$ along a path $\gamma_1$, which is generated by the Hamiltonian flows of $J_{24}^2$ inside ${\cal L}_a^{6j}$. Then it goes back to another point in ${\cal M}_0$ along a path $\gamma_2$, which is generated by the Hamiltonian flows of $J_{12}^2$ inside ${\cal L}_b^{6j}$. Finally it goes back to the starting point $z_0$ along a path $\gamma_3$ through an overall rotation around $-{\mathbf J}_5$ by an angle equal to twice the internal dihedral angle $\phi_5$ of the tetrahedron in Fig.\ \ref{SLQN9j: fig_tetrahedra}. We can break this loop up into two paths $\gamma^a = \gamma_1$ and $\gamma^b = -\gamma_3 - \gamma_2$ that goes from $z_0$ to $z_1$ along the two different Lagrangian manifolds.  

By choosing $z_0$ as the reference point for both actions $S_a^{6j}$ and $S_b^{6j}$, we have 

\begin{equation}
S_a^{6j}(z_0) - S_b^{6j}(z_0) = 0 \, . 
\end{equation}
We quote the results of Eq.\ (62) in  \cite{littlejohn2010} for the relative action at the other stationary phase point. It is given by

\begin{equation}
S_a^{6j}(z_1) - S_b^{6j}(z_1) = - 2  \sum_i J_i \psi_i + 2 n \pi \, ,
\end{equation}
where $\psi_i$ are the external dihedral angles of the tetrahedron in Fig.\ \ref{SLQN9j: fig_tetrahedra}, and the sum is over $i = 1,2,4,5, 12, 24$, and $n$ is some integer to contributes to an overall phase. The amplitudes at both stationary point are given by
\begin{equation}
\Omega_{0, \, 1} = \frac{1}{2 \sqrt{12 \pi V}} \, ,
\end{equation}
where $V$ is the volume of the tetrahedron in Fig.\ \ref{SLQN9j: fig_tetrahedra}. Altogether,  Eq.\  (\ref{SLQN9j: eq_general_formula}) becomes

\begin{eqnarray}
\label{SLQN9j: eq_9j_formula_2}
&& \braket{b | a}   \\  \nonumber
&&= \frac{e^{i \kappa} }{2 \sqrt{12 \pi V}} \left\{  (\tau^b(z_0))^\dagger \tau^a(z_0)   \right. \\  \nonumber 
&&\left. + \,  \left( U_b \tau^b(z_0) \right)^\dagger \left( U_a \tau^a(z_0) \right) 
 \text{exp} \left[ -  2i  \sum_i J_i \psi_i - \frac{i \mu_1 \pi}{2} \right]    \right\}  \, ,   
\end{eqnarray}
where $e^{i \kappa}$ is some overall phase, and $U_a$, $U_b$ are rotation matrices associated with $\gamma^a$ and $\gamma^b$, respectively.

\subsection{The reference point $z_0$ and the spinor products}

\begin{figure}[tbhp]
\begin{center}
\includegraphics[width=0.40\textwidth]{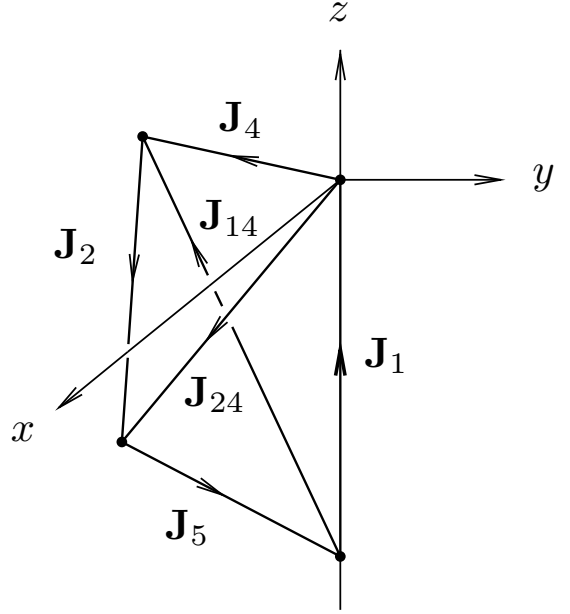}
\caption{The configuration corresponding to the reference point $z_0$. }
\label{SLQN9j: fig_z_0_config}
\end{center}
\end{figure}

We choose the vector configurations associated with $z_0$ to have a negative signed volume $V$, where $6 V = {\bf J}_1 \cdot ({\bf J}_2 \times {\bf J}_4)$. This choice of the starting point is in agreement with the choice made in \cite{littlejohn2010}. In addition, we choose a particular orientation for the starting point $z_0$, as follows. We put ${\mathbf J}_1$ along the $z$-axis, and ${\mathbf J}_1 \times {\mathbf J}_{24}$ along the $y$-axis, so that ${\mathbf J}_{24}$ and ${\mathbf J}_5$ lie inside the $xz$-plane, as illustrated in Fig.\ \ref{SLQN9j: fig_z_0_config}. Let the  the inclination and azimuth angles $(\theta, \phi)$ denote the direction of the vector ${\mathbf J}_4$. From Fig.\ \ref{SLQN9j: fig_z_0_config}, we see that $| \phi |$ is the angle between the $({\mathbf J}_1, {\mathbf J}_{24})$ plane and the $({\mathbf J}_1, {\mathbf J}_4)$ plane, so $\phi = - \phi_1$, which is defined in Eq.\  (\ref{SLQN9j: eq_phi1_def}). Also see Fig.\ \ref{SLQN9j: fig_tetrahedra_1p}. The inclination angle is simply the angle $\theta$ between ${\mathbf J}_1$ and ${\mathbf J}_4$, so it is the same $\theta$ defined in  Eq.\  (\ref{SLQN9j: eq_theta_def}).

The gauge choices for the spinors at the reference point are arbitrary, and they contribute only an overall phase. To be concrete, we choose the spinor $\tau^a$ at $z_0$ to be the standard eigenvector for $S_z$, that is,  

\begin{equation}
\tau_\alpha^a(z_0)  = \delta_{\alpha \mu} \, .
\end{equation}
For the spinor $\tau^b$ at $z_0$, we choose it to be an eigenvector of ${\mathbf J}_4 \cdot {\mathbf S}$ in the north standard gauge, 

\begin{equation}
\tau_\alpha^b(z_0) = e^{- i (\alpha - \nu) \phi_1} \, d^s_{\nu \alpha } (\theta) \, . 
\end{equation}

Taking the spinor inner product, we obtain  

\begin{equation}
(\tau^b(z_0))^\dagger (\tau^a(z_0))  = e^{i (\mu - \nu) \phi_1} \, d^s_{\nu \mu} (\theta) \, .
\end{equation}
To evaluate the other spinor product, we need to first find the rotation matrices $U_a$ and $U_b$, which are generated from $\gamma_a$ and $\gamma_b$, respectively. The path $\gamma_a = \gamma_1$ has no flows generated by the total angular momentum, so 

\begin{equation}
U_a = 1 \, . 
\end{equation}
The path $\gamma_b = - \gamma_3 - \gamma_2$ contains only an overall rotation around ${\mathbf J}_5$,  so 

\begin{equation}
U_b = U( \hat{\mathbf J}_5, 2 \phi_5) \, . 
\end{equation}
 
\begin{figure}[tbhp]
\begin{center}
\includegraphics[width=0.32\textwidth]{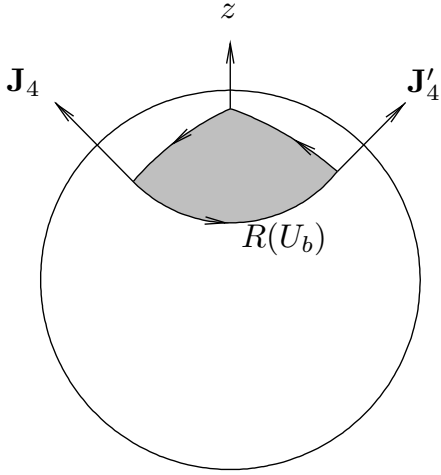}
\caption{The holonomy $H_4$ is the area bounded by the curve from $\hat{z}$ to $\hat{j}_4$ along a great circle, and the curve from $U_b$ that rotates $\hat{j}_4$ to $\hat{j}_4'$, and the curve rotating $\hat{j}_4'$ back to $\hat{z}$ along another great circle. }
\label{SLQN9j: holonomy_area}
\end{center}
\end{figure}

The rotation $U_b$ effectively moves ${\mathbf J}_4$ to its mirror image ${\mathbf J}_4'$ across the $xz$-plane, which has the direction given by  $(\theta, \phi_1)$. Thus $U_b \, \tau^b(z_0)$ is an eigenvector of ${\mathbf J}_4' \cdot {\mathbf S}$, and is up to a phase equal to the eigenvector of ${\mathbf J}_4' \cdot {\mathbf S}$ in the north standard gauge. We have

\begin{equation}
[U_b \, \tau^b(z_0)]_\alpha = e^{i \nu H_4}  \, e^{ i (\alpha - \nu) \phi_1} \, d^s_{\nu \alpha } (\theta)  \, ,
\end{equation}
where $H_4$ in the phase factor is equal to the area of a spherical triangle on a unit sphere. See Fig.\ \ref{SLQN9j: holonomy_area}. 

Therefore, the spinor inner product at the intersection ${\cal M}_1$ is 

\begin{equation}
\left(U_b \tau^b(z_0)\right)^\dagger  \left(U_a \tau^a(z_0)\right) 
= e^{i \nu H_4} \, e^{-  i (\mu - \nu) \phi_1} \, d^s_{\nu \mu } (\theta)  \, .
\end{equation}
 
Putting this back into  Eq.\  (\ref{SLQN9j: eq_9j_formula_2}), we have an asymptotic formula for the inner product,
\begin{eqnarray}
\label{SLQN9j: eq_9j_formula_3a}
&& \braket{b | a}  
= \frac{e^{i\kappa_1} }{\sqrt{12 \pi V}} \, d^s_{\nu \mu} (\theta)   \\  \nonumber
&& \cos \left[- \sum_i J_i \psi_i  - \frac{\mu_1 \pi}{4} - \mu \phi_1 + \nu \left( \frac{H_4}{2} + \phi_1 \right)  \right] \, .
\end{eqnarray}

Using a different choice of the reference point and paths, we can derive an alternative expression for the inner product, and eliminate the term $H_4$. Let us choose a new reference point $z_0$ to correspond to an orientation where ${\mathbf J}_4$ is along the $z$-axis, and ${\mathbf J}_4 \times {\mathbf J}_{12}$ along the $y$-axis. For the paths, we choose $\gamma^a = \gamma_3 + \gamma_1$ and $\gamma^b = - \gamma_2$. Through essentially the same arguments, we find 

\begin{eqnarray}
\label{SLQN9j: eq_9j_formula_3b}
&& \braket{b | a}  
= \frac{e^{i \kappa_4} }{\sqrt{12 \pi V}} \, d^s_{\nu \mu} (\theta)   \\  \nonumber
&& \cos \left[ - \sum_i J_i \psi_i - \frac{\mu_1 \pi}{4} +  \mu \left(\frac{H_1}{2} + \phi_4 \right) - \nu \phi_4)  \right] \, ,
\end{eqnarray}
where $H_1$ is another holonomy for the ${\mathbf J}_1$ vector.

Because the quantities $\psi_i,  \phi_1, \phi_4, H_1, H_4$ depend only on the geometry of the tetrahedron, and are independent of $\mu$ and $\nu$, we conclude that the argument in the cosine must be linear in $\mu$ and $\nu$. Equating the two arguments of the cosine in Eq.\ (\ref{SLQN9j: eq_9j_formula_3a}) and in Eq.\ (\ref{SLQN9j: eq_9j_formula_3b}), we find the linear term to be $- ( \mu \phi_1 + \nu \phi_4 )$. We find that Eq.\ (\ref{SLQN9j: eq_9j_formula_3a}) and Eq.\ (\ref{SLQN9j: eq_9j_formula_3b}) become

\begin{eqnarray}
\label{SLQN9j: eq_main_formula_wo_phase}
\braket{b | a}  
&=& \frac{e^{i\kappa'} }{\sqrt{12 \pi V}} \, d^s_{\nu \mu} (\theta)   \\  \nonumber
&& \cos \left( \sum_i J_i \psi_i + \frac{ \mu_1 \pi}{4} +  \mu \phi_1 + \nu \phi_4 \right) \, . 
\end{eqnarray}
Using formulas  (10.9.2) and (10.9.6) in \cite{varshalovich1981} for the special cases of the $9j$-symbol for $s=0$ and $s=1$, and through numerical experimentation, we have determined the overall phase to be

\begin{equation}
e^{i\kappa'} = (-1)^{j_1+j_2+j_4+j_5+2s+\nu}  \, ,  
\end{equation}
where the factor $(-1)^\nu$ is necessary to make the expression Eq.\  (\ref{SLQN9j: eq_main_formula_wo_phase}) symmetrical between $\mu$ and $\nu$, because $d^s_{\nu\mu}(\theta) = (-1)^{\mu - \nu} d^s_{\mu\nu}(\theta)$. We see that the formula Eq.\  (\ref{SLQN9j: eq_main_formula}) satisfies the symmetry of the $9j$-symbol obtained from a reflection of the $j$'s across the diagonal, which corresponds to swapping the states $\ket{a}$ and $\ket{b}$. 

Again, through numerical experimentation, we found the Maslov index to be

\begin{equation}
\mu_1 = 1 - 4s \, . 
\end{equation}
Altogether, we have derived a new asymptotic formula Eq.\  (\ref{SLQN9j: eq_main_formula}) for the $9j$-symbol, which we repeat here: 

\begin{eqnarray}
&& \left\{
  \begin{array}{ccc}
    j_1 & j_2 & j_{12} \\ 
    s & j_4 & j_{34} \\ 
    j_{13} & j_{24} & j_5 \\
  \end{array} 
  \right\}
	= \frac{(-1)^{j_1+j_2+j_4+j_5+2 s +\nu}}{\sqrt{(2j_{13}+1) (2j_{34}+1)  \, (12 \pi V)}} \,  \nonumber   \\  
&&	\cos \left(  \sum_i  \, J_i \, \psi_i  + \frac{\pi}{4} - s \pi + \mu \phi_1 +\nu \phi_4  \right) 
	\, d^{s}_{\nu \, \mu} (\theta) \, . 
	\label{SLQN9j: eq_main_formula1}
\end{eqnarray}

\section{\label{SLQN9j: sec_numeric} PLOTS}

\begin{figure}[tbhp]
\begin{center}
\includegraphics[width=0.45\textwidth]{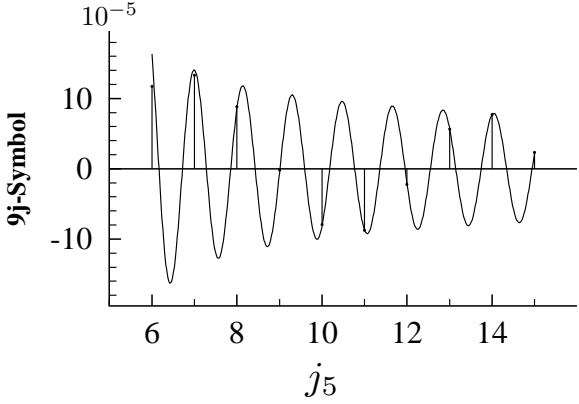}
\caption{Comparison of the exact $9j$-symbol (vertical sticks and dots) and the asymptotic formula Eq.\  (\ref{SLQN9j: eq_main_formula1}) in the classically allowed region, for the values of $j$'s shown in  Eq.\  (\ref{SLQN9j: eq_9j_values_case1}). }
\label{SLQN9j: fig_plot_9j_case1}
\end{center}
\end{figure}

\begin{figure}[tbhp]
\begin{center}
\includegraphics[width=0.45\textwidth]{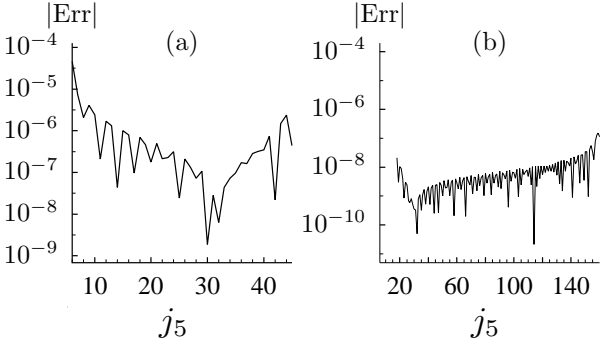}
\caption{Absolute value of the error of the asymptotic formula Eq.\  (\ref{SLQN9j: eq_main_formula1}) for (a) the case in Eq.\  (\ref{SLQN9j: eq_9j_values_case1}), and (b) the case in Eq.\  (\ref{SLQN9j: eq_9j_values_case2}). The error is defined as the difference between the approximate value and the exact value. }
\label{SLQN9j: fig_plot_9j_error}
\end{center}
\end{figure}

We illustrate the accuracy of our formula by plotting our approximation Eq.\  (\ref{SLQN9j: eq_main_formula1}) against the exact $9j$-symbols in the classical region for the following values of the $j$'s:

\begin{equation}
\label{SLQN9j: eq_9j_values_case1}
\left\{
  \begin{array}{ccc}
    j_1 & j_2 & j_{12} \\ 
    s & j_4 & j_{34} \\ 
    j_{13} & j_{24} & j_5 \\
  \end{array} 
  \right\}
	= 
	\left\{
  \begin{array}{rrr}
    51/2 & 53/2 & 28 \\ 
    1/2 & 47/2 & 24 \\ 
    25 & 27 & j_5 \\
  \end{array} 
  \right\} \, . 
\end{equation}
The result is shown in Fig.\ \ref{SLQN9j: fig_plot_9j_case1}. We see that our approximations are in good agreement with the exact values of the $9j$-symbol in the classically allowed region, away from the caustic. However, it gets progressively worse as we come closer to the caustic, where $V = 0$ and the amplitude blows up. We can see this clearly from the error plot in part (a) of Fig.\ \ref{SLQN9j: fig_plot_9j_error}.

\begin{figure}[tbhp]
\begin{center}
\includegraphics[width=0.45\textwidth]{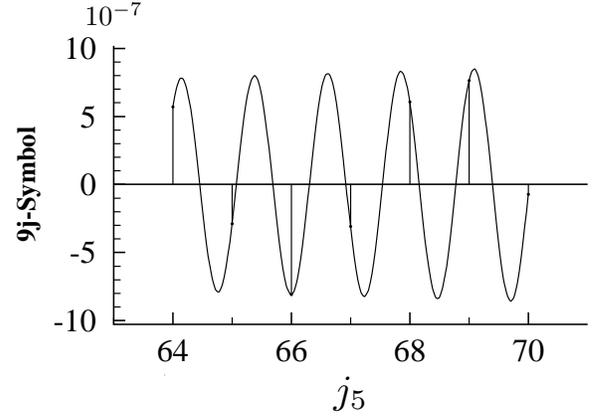}
\caption{Comparison of the exact $9j$-symbol (vertical sticks and dots) and the asymptotic formula Eq.\  (\ref{SLQN9j: eq_main_formula1}) in the classically allowed region away from the caustics, for the values of $j$'s shown in  Eq.\  (\ref{SLQN9j: eq_9j_values_case2}). }
\label{SLQN9j: fig_plot_9j_classical_region}
\end{center}
\end{figure}

\begin{figure}[tbhp]
\begin{center}
\includegraphics[width=0.45\textwidth]{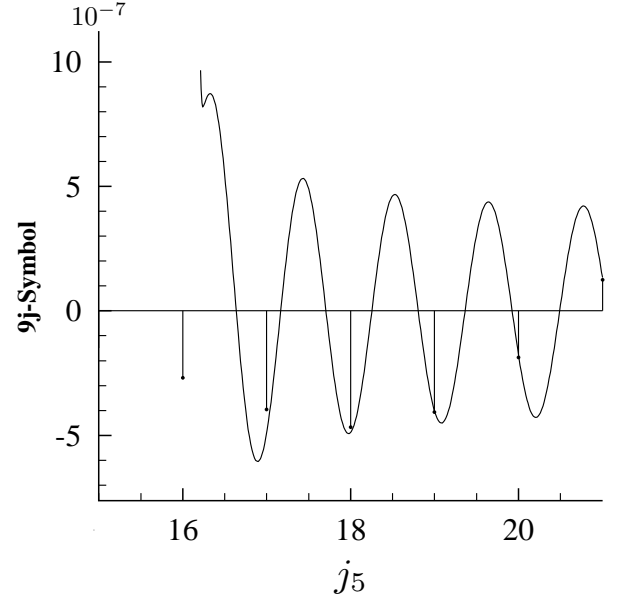}
\caption{Comparison of the exact $9j$-symbol (vertical sticks and dots) and the asymptotic formula Eq.\  (\ref{SLQN9j: eq_main_formula1}) near a caustic, for the values of $j$'s shown in  Eq.\  (\ref{SLQN9j: eq_9j_values_case2}).}
\label{SLQN9j: fig_plot_9j_caustic_region}
\end{center}
\end{figure}

Since the asymptotic formula Eq.\  (\ref{SLQN9j: eq_main_formula1}) should become more accurate as the values of the $j$'s get larger, we plot the formula against the exact $9j$ for another example Eq.\  (\ref{SLQN9j: eq_9j_values_case2}), which is roughly Eq.\  (\ref{SLQN9j: eq_9j_values_case1}) rescaled by a factor of $4$. The result in the classically allowed region away from the caustic is shown in Fig.\ \ref{SLQN9j: fig_plot_9j_classical_region}, and the result in regions close to the caustic is shown in Fig.\ \ref{SLQN9j: fig_plot_9j_caustic_region}. The errors for this case is shown in part (b) of Fig.\ \ref{SLQN9j: fig_plot_9j_error}. We can see that the error also scales with the values of the $j$'s. 

\begin{equation}
\label{SLQN9j: eq_9j_values_case2}
\left\{
  \begin{array}{ccc}
    j_1 & j_2 & j_{12} \\ 
    s & j_4 & j_{34} \\ 
    j_{13} & j_{24} & j_5 \\
  \end{array} 
  \right\}
	= 
	\left\{
  \begin{array}{rrr}
    201/2 & 205/2 & 89 \\ 
    3/2 & 197/2 & 99 \\ 
    100 & 92 & j_5 \\
  \end{array} 
  \right\} \, . 
\end{equation}

\section{COMMENTS AND CONCLUSIONS}

We have computed the asymptotic formula for the Wigner $9j$-symbol when some angular momenta are large while others are small. In so doing, we have formulated a gauge-invariant form of the multicomponent WKB wave-functions that appears in the definition of the $9j$-symbol. Our gauge-invariant form retains the factorization of the wavefunction into a scalar WKB part and a spinor part. Currently, there is no general theory of finding gauge-invariant multicomponent WKB wave-functions, so our formulation here provides an interesting special case for investigating when such a gauge-invariant form is possible. 

There are several directions to extend the present work. A natural extension is to find the analytic continuation of our formula to the classically forbidden region where the Lagrangian manifolds do not have real intersections. While the analytic continuation of the cosine term in the $6j$-symbol into its classically forbidden region is known, it is unclear what we should do with the $d$-matrix and the complex argument $\theta$ in the forbidden region. This will be the subject of future research. 

Another extension is to apply our technique to the $12j$-symbol with two small angular momenta, and to the $15j$-symbol with three small angular momenta. This is easy, because the same Lagrangian manifolds in this paper, namely, those that appear in the semiclassical analysis of the $6j$-symbol in the body of this paper, also appear in those cases for the $12j$- and $15j$-symbols. Since the derivations are almost identical, we will simply display these results for the $12j$-symbol in Appendix \ref{appendix_12j}, and results for the $15j$-symbol in Appendix \ref{appendix_15j}.

It is more difficult to apply our method to the case of the $12j$-symbol with one small angular momentum, or to the $15j$-symbol with two small angular momenta. The relevant Lagrangian manifolds are those that appear in the semiclassical analysis of the $9j$-symbol. We will present our results for those cases in a future article.

One limitation of our method is that it is not applicable to cases where there are three small angular momenta in a single row or column of the $9j$-symbol. The asymptotic formula for the $(6, 3)$ case when the three small angular momenta are placed in the same row is still unknown. In such cases, the matrix operators that define the quantum states are more complicated. We expect a solution  will involve a combination of the method developed in this paper and the summation formula used in \cite{anderson2008, anderson2009, watson1999b, watson1999a}. This interplay of two very different techniques will be interesting for a future research project.

\appendix

\section{\label{appendix_9j} The $9j$-symbol with two small angular momenta}

\begin{figure}[tbhp]
\begin{center}
\includegraphics[width=0.35\textwidth]{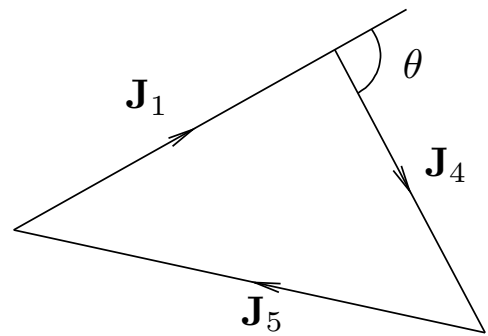}
\caption{The angle $\theta$ is defined as the exterior angle between the edges $J_1$ and $J_4$ in a triangle having the three edge lengths $J_1, J_4, J_5$.}
\label{SLQN9j: fig_plot_9j_2_7_triangle}
\end{center}
\end{figure}

We can also take the two angular momenta to be small. Let us write down the definition of the $9j$-symbol when $s_2=j_2$ and $s_3=j_3$ are small. It is

\begin{eqnarray}
 && \left\{
   \begin{array}{ccc}
    j_1 & s_2 & j_{12} \\ 
    s_3 & j_4 & j_{34} \\ 
    j_{13} & j_{24} & j_5 \\
  \end{array} 
  \right \}    \\  \nonumber
&=& \frac{\braket{ b  |  a } }{[(2j_{12}+1)(2j_{34}+1)(2j_{13}+1)(2j_{24}+1)]^{\frac{1}{2}}} \, , 
\end{eqnarray}
where 
	
\begin{equation}
\ket{a} =  \left| 
\begin{array} { @{\;}c@{\;}c@{\;}c@{\;}c@{\,}c@{\;}c@{\;}c@{\;}c@{}}
	\hat{I}_1 & {\mathbf S}_2^2 & {\mathbf S}_3^2 & \hat{I}_4 & \hat{I}_5 & \hat{\mathbf J}_{13}^2 & \hat{\mathbf J}_{24}^2 & \hat{\mathbf J}_{\text{tot}}  \\
	j_1 & s_2 & s_3 & j_4 & j_5 & j_{13} & j_{24} & {\mathbf 0} 
\end{array}  \right> \, , 
\end{equation}

\begin{equation}
\ket{b} =  \left| 
\begin{array} { @{\;}c@{\;}c@{\;}c@{\;}c@{\,}c@{\;}c@{\;}c@{\;}c@{}}
	\hat{I}_1 &  {\mathbf S}_2^2  & {\mathbf S}_3^2 & \hat{I}_4 & \hat{I}_5 & \hat{\mathbf J}_{12}^2 & \hat{\mathbf J}_{34}^2 & \hat{\mathbf J}_{\text{tot}}  \\
	j_1 & s_2 & s_3 & j_4 & j_5 & j_{12} & j_{34} & {\mathbf 0} 
\end{array}  \right> \, . 
\end{equation}
We choose the four spinors just as in the $9j$ case with one small angular momentum. Then the Hamilton-Jacobi equations for the operators $\hat{J}_{13}^2, \hat{J}_{24}^2, \hat{J}_{12}^2$, and $\hat{J}_{34}^2$, simply pick out the polarizations. The remaining Hamilton-Jacobi equations for both states after the perturbations are performed become

\begin{eqnarray*}
J_1 = (j_1+1/2)  \hbar \,  ,  \\
J_4 = (j_4+1/2)  \hbar \,  ,  \\
J_5 = (j_5+1/2)  \hbar \,  ,  \\
{\mathbf 0} ={\mathbf J}_1 +{\mathbf J}_4 +{\mathbf J}_5 \,  .
\end{eqnarray*}
These equations describe a Lagrangian manifold that corresponds to a triangle having the three edge lengths $J_1, J_4, J_5$, illustrated in Fig.\ \ref{SLQN9j: fig_plot_9j_2_7_triangle}. 

Because the two Lagrangian manifolds are identical, the inner product of the scalar part of the WKB wave-functions is unity, and only the spinor products contribute. After putting in the correct overall phase, we get

\begin{eqnarray}
\label{SLQN9j: eq_9j_2_7_formula}
&& \left\{
  \begin{array}{ccc}
    j_1 & s_2 & j_{12} \\ 
    s_3 & j_4 & j_{34} \\ 
    j_{13} & j_{24} & j_5 \\
  \end{array} 
  \right\}  \\   \nonumber
&=& \frac{(-1)^{2(j_4+j_5)+s_2+s_3+j_{12}+j_{13}}}{\sqrt{(2j_{13}+1)(2j_{24}+1)(2j_{12}+1)(2j_{34}+1)}}  \\   \nonumber
&& \;  d^{(s_2)}_{j_{12}-j_1 , \, j_{24}-j_4} (\theta) \, d^{(s_3)}_{j_{13}-j_1  , \, j_{34}-j_4} (\theta)  \, ,  
\end{eqnarray}
where $\theta$ is the exterior angle between the vectors $\vec{J}_1$ and $\vec{J}_4$ in the triangle formed by $J_1$, $J_4$, and $J_5$, as illustrated in Fig.\ \ref{SLQN9j: fig_plot_9j_2_7_triangle}. The angle $\theta$ is given by

\begin{equation}
	\theta = \cos^{-1} \left( \frac{{\mathbf J}_1 \cdot {\mathbf J}_4}{J_1 \, J_4} \right) = \pi - \cos^{-1} \left( \frac{J_1^2 + J_4^2 - J_5^2 }{2 \, J_1 \, J_4} \right) \, .
\end{equation}

\begin{figure}[tbhp]
\begin{center}
\includegraphics[width=0.45\textwidth]{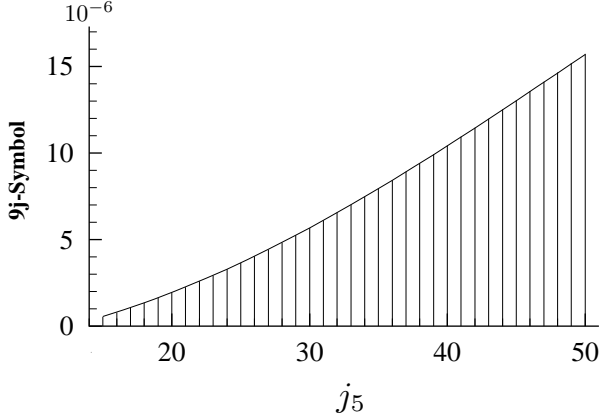}
\caption{Comparison of exact $9j$-symbol (vertical sticks) and the asymptotic formula Eq.\  (\ref{SLQN9j: eq_9j_2_7_formula}), for the values of $j$'s shown in  Eq.\  (\ref{9j_2_7_values}).}
\label{SLQN9j: fig_plot_9j_2_7_case}
\end{center}
\end{figure}

We plot the formula Eq.\  (\ref{SLQN9j: eq_9j_2_7_formula}) below against exact values of the $9j$-symbol in Fig.\ \ref{SLQN9j: fig_plot_9j_2_7_case}, for the following values of the $j$'s:

\begin{equation}
\label{9j_2_7_values}
\left\{
  \begin{array}{ccc}
    j_1 & s_2 & j_{12} \\ 
    s_3 & j_4 & j_{34} \\ 
    j_{13} & j_{24} & j_5 \\
  \end{array} 
  \right\}
= \left\{
  \begin{array}{ccc}
    67 & 1/2 & 135/2 \\ 
    3/2 & 54 & 111/2 \\ 
    135/2 & 107/2 & j_5 \\
  \end{array} 
  \right\} \, . 
\end{equation}

\pagebreak

\section{\label{appendix_12j}  The $12j$-symbol with two small angular momenta}

We now treat the case where two angular momenta are small. We take $j_1 = s_1$ and $j_5=s_5$ to be small. Using (A4) in \cite{jahn1954} for the definition for the $12j$-symbol of the first kind, the $12j$-symbol can again be written as a scalar product of two multicomponent wave-functions, as follows: 

\begin{equation}
\left\{
   \begin{array}{cccc}
    s_1 & j_2 & j_{12} & j_{125} \\ 
    j_3 & j_4 & j_{34} &  j_{135}  \\ 
    j_{13} & j_{24} & s_5 & j_6  \\
  \end{array} 
  \right \}    
= \frac{\braket{ b  |  a } }{ \{ [j_{12}][j_{34}][j_{13}][j_{24}] [j_{125}] [j_{135}] \}^{\frac{1}{2}}}  \, ,
\end{equation}
where

\begin{equation}
\ket{a} =  \left| 
\begin{array} { @{\;}c@{\,}c@{\,}c@{\,}c@{\;}c@{\;}c@{\;}c@{\;}c@{\;}c@{\;}c@{}}
	{\bf S}_1^2 & \hat{I}_2 & \hat{I}_3  & \hat{I}_4 &  {\bf S}_5^2 & \hat{I}_6 & \hat{\bf J}_{12}^2 & \hat{\bf J}_{34}^2 & \hat{\bf J}_{125}^2 &  \hat{\bf J}_{\text{tot}}  \\
	s_1 & j_2 & j_3 & j_4 & s_5 & j_6 & j_{12} & j_{34} & j_{125} & {\bf 0} 
\end{array}  \right> \,  , 
\end{equation}

\begin{equation}
\ket{b} =  \left| 
\begin{array} { @{\;}c@{\,}c@{\,}c@{\,}c@{\;}c@{\;}c@{\;}c@{\;}c@{\;}c@{\;}c@{}}
	{\bf S}_1^2 & \hat{I}_2 & \hat{I}_3  & \hat{I}_4 &  {\bf S}_5^2 & \hat{I}_6 & \hat{\bf J}_{13}^2 & \hat{\bf J}_{24}^2 & \hat{\bf J}_{135}^2 &  \hat{\bf J}_{\text{tot}}  \\
	s_1 & j_2 & j_3 & j_4 & s_5 & j_6 & j_{13} & j_{24} & j_{135} & {\bf 0} 
\end{array}  \right>  \, . 
\end{equation}
Using the formalism developed in this paper, we can now find a new asymptotic formula for the $12j$-symbol up to an overall phase. By expressing the $12j$-symbol in terms of the $9j$-symbol in the special cases $s_1 = 0$ or $s_5=0$ from (A1) in \cite{jahn1954}, and by using Eq.\  (\ref{SLQN9j: eq_main_formula}), we can determine the overall phase. The result is an asymptotic formula for the $12j$-symbol when two of the angular momenta are small: 

\begin{widetext}
\begin{eqnarray}
\label{SLQN9j: eq_formula_12j_2_10}
 \left\{
  \begin{array}{cccc}
    s_1 & j_2 & j_{12} & j_{125} \\ 
    j_3 & j_4 & j_{34} &  j_{135}  \\ 
    j_{13} & j_{24} & s_5 & j_6  \\
  \end{array} 
  \right\}    
&=& (-1)^{ j_{24}+j_{34}+j_{125}+j_{135}+j_1+j_5+\nu_1 + \nu_5} \; \frac{ d^{s_1}_{\nu_1 \, \mu_1} (\theta) \; d^{s_5}_{\nu_5 \, \mu_5} (\theta)}{\sqrt{ [j_{12}][j_{125}][j_{13}][j_{135}] (12 \pi V) }} \,    \\   \nonumber
&& \cos \left(  \sum_i  \, (j_i+\frac{1}{2}) \, \psi_i  + \frac{\pi}{4} - ( s_1 + s_5) \pi + (\mu_1+\mu_5) \phi_2 + (\nu_1 + \nu_5) \phi_3  \right)   \, .
\end{eqnarray}
\end{widetext}

\begin{figure}[tbhp]
\begin{center}
\includegraphics[width=0.35\textwidth]{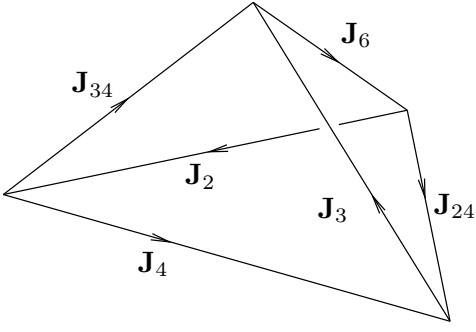}
\caption{The volume $V$ and the external dihedral angles $\psi_i$ are defined on the tetrahedron with the six edge lengths $J_3, J_5, J_2, J_6, J_{24}, J_{34}$. }
\label{SLQN9j: fig_tetrahedra_12j}
\end{center}
\end{figure}

\begin{figure}[tbhp]
\begin{center}
\includegraphics[width=0.48\textwidth]{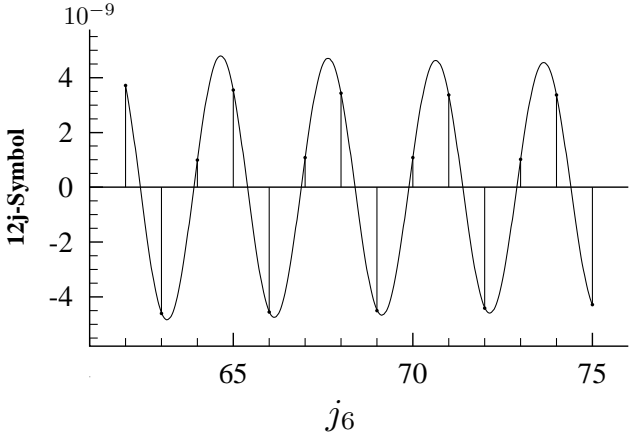}
\caption{Comparison of the exact $12j$-symbol (vertical sticks and dots) and the asymptotic formula Eq.\  (\ref{SLQN9j: eq_formula_12j_2_10}), in the classically allowed region away from the caustics. The values used are those in  Eq.\  (\ref{SLQN9j: eq_12j_values_2_10}). }
\label{SLQN9j: fig_plot_12j_2_10}
\end{center}
\end{figure}

Here $\mu_1 = j_{12}-j_2$, $\nu_1 = j_{13}-j_3$, $\mu_5 = j_{125}-j_{12}$, and $\nu_5 = j_{135}-j_{13}$. The sum in the argument of the cosine runs over the six large angular momenta $i = 2, 3, 5, 6, 24, 34$. The geometric quantities $V$, $\psi_i$, $\phi_2$, $\phi_3$, and $\theta$ are related to the tetrahedron in Fig.\ \ref{SLQN9j: fig_tetrahedra_12j}, which has the six edge lengths $J_i = j_i + 1/2$, $i = 2,3,5,6,24,34$. As before, $V$ is the volume, and each $\psi_i$ is the external dihedral angle at the edge $J_i$. The angle $\phi_2$ is the angle between the plane spanned by $({\bf J}_2, {\bf J}_3)$ and the plane spanned by $({\bf J}_2, {\bf J}_{6})$.  The angle $\phi_3$ is the angle between the plane defined by $({\bf J}_2, {\bf J}_3)$ and the plane defined by $({\bf J}_3, {\bf J}_{6})$. The angle $\theta$ is the angle between ${\bf J}_2$ and ${\bf J}_3$. The explicit expression for the angles $\phi_{2}$, $\phi_{3}$, and $\theta$ are given by the following equations:

\begin{eqnarray}
\label{SLQN9j: eq_phi_2_def}
\phi_{2} &=&  \pi - \cos^{-1} \left( \frac{ ({\bf J}_{2} \times {\bf J}_{3} ) \cdot ({\bf J}_{2} \times {\bf J}_{6} ) }{ | {\bf J}_{2} \times {\bf J}_{3} | \,  | {\bf J}_{2} \times {\bf J}_{6}  |} \right) \, ,    \\
\label{SLQN9j: eq_phi_3_def}
\phi_{3} &=& \pi -  \cos^{-1} \left(  \frac{ ({\bf J}_{3} \times {\bf J}_{2} ) \cdot ({\bf J}_{3} \times {\bf J}_{6} ) }{ | {\bf J}_{3} \times {\bf J}_{2} | \,  | {\bf J}_{3} \times {\bf J}_{6}  |}  \right)  \, ,   \\
\label{SLQN9j: eq_theta_12_3_9_def}
\theta &=& \cos^{-1} \left( \frac{ {\bf J}_{2} \cdot {\bf J}_{3} }{J_{2} J_{3} }  \right)  \, . 
\end{eqnarray}

We illustrate the accuracy of the approximation Eq.\  (\ref{SLQN9j: eq_formula_12j_2_10}) by plotting it against the exact $12j$-symbol in Fig.\ \ref{SLQN9j: fig_plot_12j_2_10}  for the following values of the $j$'s:  

\begin{equation}
\label{SLQN9j: eq_12j_values_2_10}
\left\{
  \begin{array}{cccc}
    s_1 & j_2 & j_{12} & j_{125} \\ 
    j_3 & j_4 & j_{34} &  j_{135}  \\ 
    j_{13} & j_{24} & s_5 & j_6  \\
  \end{array} 
  \right\}   
=
	\left\{
  \begin{array}{rrrr}
    1/2 & 201/2 & 100 & 101 \\ 
    213/2 & 199/2 & 117 & 105 \\ 
    106 & 98 & 1 & j_6\\
  \end{array} 
  \right\}  \, . 
\end{equation}
We see that the agreement is excellent.  \\

\section{\label{appendix_15j} The $15j$-symbol with three small angular momenta}

\begin{figure}[tbhp]
\begin{center}
\includegraphics[width=0.30\textwidth]{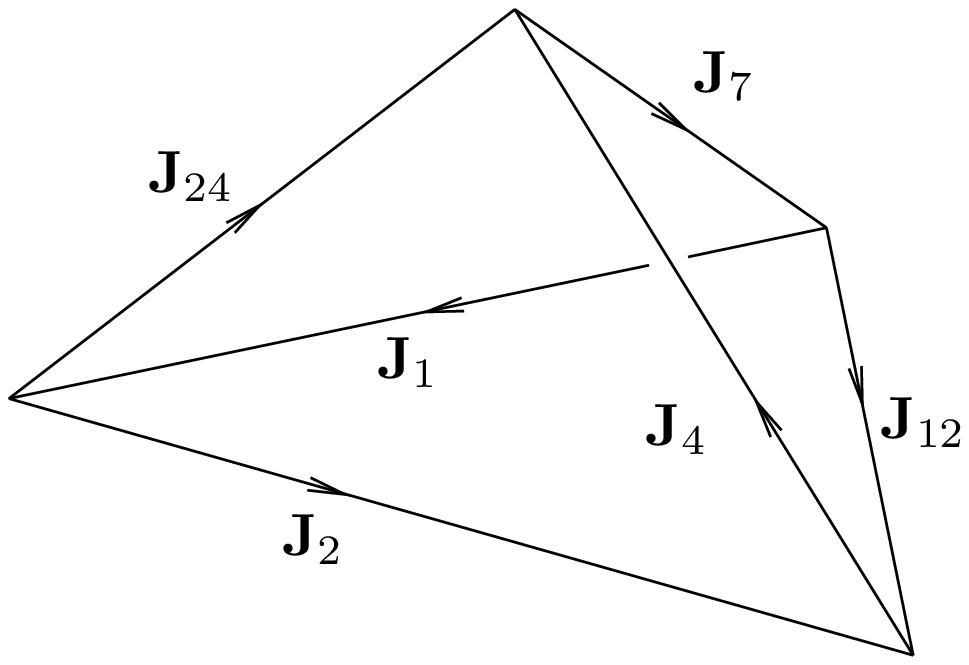}
\caption{The volume $V$ and the external dihedral angles $\psi_i$ are defined on the tetrahedron with the six edge lengths $J_1, J_2, J_4, J_7, J_{12}, J_{24}$. }
\label{SLQN9j: fig_tetrahedra_15j_3_12}
\end{center}
\end{figure}

We now treat the case where three quantum numbers are small. We take $j_3 = s_3$, $j_5=s_5$, and $j_6=s_6$ to be small. By using the spin network definition on page 66 in \cite{yutsis1962} for the definition for the $15j$-symbol of the first kind, we write the $15j$-symbol as a scalar product of two multicomponent wave-functions,

\begin{eqnarray}
&& \left\{
   \begin{array}{ccccc}
    j_1 & j_2 & j_{12} & j_{125} & j_{1256} \\ 
    s_3 & j_4 & j_{34} &  j_{135}  & j_{1356} \\ 
    j_{13} & j_{24} & s_5 & s_6  & j_7  \\
  \end{array} 
  \right \}     \\  \nonumber
&=& \frac{\braket{ b  |  a } }{ \{ [j_{12}][j_{34}][j_{13}][j_{24}] [j_{125}] [j_{135}] [j_{1256}] [j_{1356}] \}^{\frac{1}{2}}}  \, , 
\end{eqnarray}
where

\begin{equation}
\ket{a} =  \left| 
\begin{array} { @{\,}c@{\;}c@{\;}c@{\,}c@{\;}c@{\;}c@{\,}c@{\;}c@{\;}c@{\;}c@{\;}c@{\;}c@{}}
	\hat{I}_1 & \hat{I}_2 & {\bf S}_3^2 & \hat{I}_4 &  {\bf S}_5^2 & {\bf S}_6^2 &  \hat{I}_7 & \hat{\bf J}_{12}^2 & \hat{\bf J}_{34}^2 & \hat{\bf J}_{125}^2 &  \hat{\bf J}_{1256}^2 &  \hat{\bf J}_{\text{tot}}  \\
	j_1 & j_2 & s_3 & j_4 & s_5 & s_6 & j_7 & j_{12} & j_{34} & j_{125} & j_{1256} & {\bf 0} 
\end{array}  \right>  \, , 
\end{equation}

\begin{equation}
\ket{b} =  \left| 
\begin{array} { @{\,}c@{\;}c@{\;}c@{\,}c@{\;}c@{\;}c@{\,}c@{\;}c@{\;}c@{\;}c@{\;}c@{\;}c@{}}
	\hat{I}_1 & \hat{I}_2 & {\bf S}_3^2 & \hat{I}_4 &  {\bf S}_5^2 & {\bf S}_6^2 &  \hat{I}_7 & \hat{\bf J}_{13}^2 & \hat{\bf J}_{24}^2 & \hat{\bf J}_{135}^2 &  \hat{\bf J}_{1356}^2 &  \hat{\bf J}_{\text{tot}}  \\
	j_1 & j_2 & s_3 & j_4 & s_5 & s_6 & j_7 & j_{13} & j_{24} & j_{135} & j_{1356} & {\bf 0} 
\end{array}  \right>  \, . 
\end{equation}
Following the strategy in this paper, we derive an asymptotic formula for the $15j$-symbol when three angular momenta are small, up to an overall phase. The result is the following formula:

\begin{widetext}
\begin{eqnarray}
\label{SLQN9j: eq_main_formula_15j_3_12}
  \left\{
  \begin{array}{ccccc}
    j_1 & j_2 & j_{12} & j_{125} & j_{1256}\\ 
    s_3 & j_4 & j_{34} &  j_{135} & j_{1356} \\ 
    j_{13} & j_{24} & s_5 & s_6 & j_7 \\
  \end{array} 
  \right\}    
&=&  
(-1)^{ j_1 + j_2 + j_4 + j_7 + 2s_3 + \nu_3 + \mu_5 + \mu_6} \; \frac{ d^{s_3}_{\nu_3 \, \mu_3} (\theta_1) \; d^{s_5}_{\nu_5 \, \mu_5} (\theta_2) \; d^{s_6}_{\nu_6 \, \mu_6} (\theta_2)}{\sqrt{ [j_{34}][j_{13}][j_{135}] [j_{1356}][j_{125}][j_{1256}]  (12 \pi V) }} \,  \\   \nonumber
&& \quad  \cos \left(  \sum_i  \, (j_i+\frac{1}{2}) \, \psi_i  + \frac{\pi}{4} - s_3  \pi  
 + \mu_3  \phi_4'  + \nu_3  \phi_1'  - (\mu_5 + \mu_6) \phi_{12} - (\nu_5 + \nu_6) \phi_1 \right)   \, .
\end{eqnarray}
\end{widetext}
Here $\mu_3 = j_{34}-j_4$, $\nu_3 = j_{13}-j_1$, $\mu_5 = j_{125}-j_{12}$, $\nu_5 = j_{135}-j_{13}$, $\mu_6 = j_{1256} - j_{125}$, and $\nu_6 = j_{1356} - j_{135}$. The angles $\phi_1$ and $\phi_{12}$ are internal dihedral angles in Fig.\  \ref{SLQN9j: fig_tetrahedra_15j_3_12}, in other words, $\phi_1 = \pi - \psi_1$ and $\phi_{12} = \pi - \psi_{12}$.  Similar to the previous cases, the angle $\phi_1'$ is the angle between the $({\bf J}_1, {\bf J}_4)$ plane and the $({\bf J}_1, {\bf J}_{24})$ plane. The angle $\phi_4'$ is the angle between the $({\bf J}_1, {\bf J}_4)$ plane and the $({\bf J}_4, {\bf J}_{12})$ plane. Here we put primes on these angles to distinguish them from the internal dihedral angles $\phi_1$ and $\phi_4$. The angle $\theta_1$ is the angle between ${\bf J}_1$ and ${\bf J}_4$. The angle $\theta_2$ is the angle between ${\bf J}_1$ and ${\bf J}_{12}$. Explicitly, the angles $\phi_1', \phi_4'$, and $\theta$ are given by the following equations:

\begin{eqnarray}
\label{SLQN9j: eq_phi_1p_def}
\phi_{1}' &=&  \pi - \cos^{-1} \left( \frac{ ({\bf J}_{1} \times {\bf J}_{4} ) \cdot ({\bf J}_{1} \times {\bf J}_{7} ) }{ | {\bf J}_{1} \times {\bf J}_{4} | \,  | {\bf J}_{1} \times {\bf J}_{7}  |} \right) \, ,    \\
\label{SLQN9j: eq_phi_4p_def}
\phi_{4}' &=& \pi -  \cos^{-1} \left(  \frac{ ({\bf J}_{4} \times {\bf J}_{1} ) \cdot ({\bf J}_{4} \times {\bf J}_{7} ) }{ | {\bf J}_{4} \times {\bf J}_{1} | \,  | {\bf J}_{4} \times {\bf J}_{7}  |}  \right)  \, ,   \\
\label{SLQN9j: eq_theta_15_3_12_def}
\theta &=& \cos^{-1} \left( \frac{ {\bf J}_{1} \cdot {\bf J}_{4} }{J_{1} J_{4} }  \right)  \, . 
\end{eqnarray}

\begin{figure}[tbhp] 
\begin{center}
\includegraphics[width=0.48\textwidth]{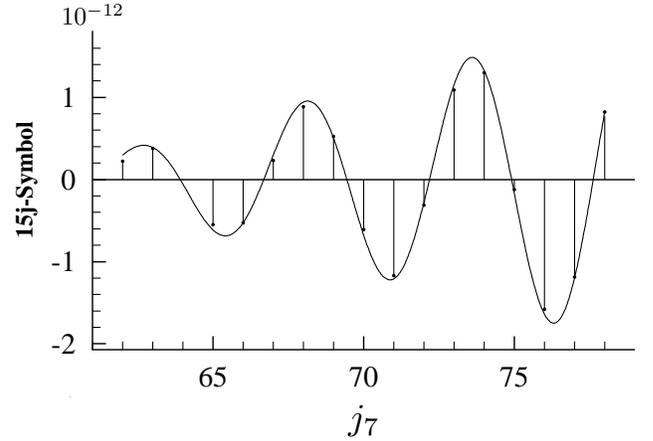}
\caption{Comparison of the exact $15j$-symbol (vertical sticks and dots) and the asymptotic formula Eq.\  (\ref{SLQN9j: eq_main_formula_15j_3_12}), for the values of $j$'s shown in Eq.\  (\ref{SLQN9j: eq_15j_values_3_12}). }
\label{SLQN9j: fig_plot_15j_3_12}
\end{center}
\end{figure}

We plot the exact values of the $15j$-symbol against our approximation Eq.\  (\ref{SLQN9j: eq_main_formula_15j_3_12}) for the following values of the $j$'s: 

\begin{eqnarray}
\label{SLQN9j: eq_15j_values_3_12}
&& \left\{
  \begin{array}{ccccc}
    j_1 & j_2 & j_{12} & j_{125} & j_{1256} \\ 
    s_3 & j_4 & j_{34} &  j_{135}  & j_{1356} \\ 
    j_{13} & j_{24} & s_5 & s_6  & j_7  \\
  \end{array} 
  \right\}     \\   \nonumber
&=& 
	\left\{
  \begin{array}{rrrrr}
    203/2 & 207/2 & 96 & 97 & 98 \\ 
     3/2 & 199/2 & 100 & 100 & 101  \\ 
    101 & 108 & 1 & 1 & j_7 \\
  \end{array} 
  \right\}  \, . 
\end{eqnarray}
We see that there are generally good agreements.

\pagebreak


\bibliography{SLQN9j}

\end{document}